\documentclass[12pt,onecolumn,peerreview]{IEEEtran}

\usepackage{fancyhdr}
\usepackage{epsfig}
\usepackage{threeparttable}
\usepackage{epsf,epsfig}
\usepackage{amsthm}
\usepackage[cmex10]{amsmath}
\usepackage{amssymb, amsfonts}
\usepackage[noadjust]{cite}
\usepackage{dsfont}
\usepackage{subfigure}
\usepackage{color}
\usepackage{soul}
\usepackage{amssymb}
 \usepackage{accents}
  \usepackage{algorithm}
   \usepackage[english]{babel}
    \usepackage{url}
    \usepackage{framed}

\newtheorem{lemma}{Lemma}

\newtheorem{proposition}{Proposition}
\newtheorem{scheme}{Scheme}   %%% DZ added
\def\qed{$\Box$}

\def\proof{\noindent{\emph{Proof:} }}

\def
\endproof{\hspace*{\fill}~\qed
\par
\endtrivlist\unskip}
\def\endproof{\hspace*{\fill}~\qed\par\endtrivlist\vskip3pt}

\def\phi{\varphi}

\def\l{\left}
\def\r{\right}
\def\({\left(}
\def\){\right)}

\def\bh{{\mathbf{h}}}

\def\bn{{\mathbf{n}}}

\def\bw{{\mathbf{w}}}
\def\bx{{\mathbf{x}}}
\def\by{{\mathbf{y}}}
\def\bz{{\mathbf{z}}}
\def\b0{{\mathbf{0}}}

\def\bF{{\mathbf{F}}}

\def\bH{{\mathbf{H}}}
\def\bI{{\mathbf{I}}}

\newtheorem {Remark}{Remark}

\renewcommand{\baselinestretch}{1.4}

\begin{document}	
\title{ Broadband Analog Aggregation for Low-Latency Federated Edge Learning (Extended Version)}
\author{Guangxu Zhu, Yong Wang and Kaibin Huang     \thanks{\setlength{\baselineskip}{13pt} \noindent G. Zhu, Y. Wong and K. Huang are with the Dept. of Electrical and Electronic Engineering at The  University of  Hong Kong, Hong Kong (Email: gxzhu@eee.hku.hk, wangyong@eee.hku.hk, huangkb@eee.hku.hk). Corresponding author: K. Huang. }}
\maketitle

%Update on Dec. 23:

%\vspace{-8mm}
\begin{abstract}
To leverage rich data distributed at the network edge, a new machine-learning paradigm, called edge learning, has emerged where learning algorithms are deployed at the edge for providing intelligent services to mobile users. While computing speeds are advancing rapidly, the communication latency is becoming the bottleneck of fast edge learning. To address this issue, this work is focused on designing a low-latency multi-access scheme for edge learning. To this end, we consider a popular privacy-preserving framework, \emph{federated edge learning} (FEEL), where a global AI-model at an edge-server is updated by aggregating (averaging) local models trained at edge devices. It is proposed that the updates simultaneously transmitted by devices over broadband channels should be analog aggregated ``over-the-air'' by exploiting the waveform-superposition property of a multi-access channel. Such \emph{broadband analog aggregation} (BAA) results in dramatical communication-latency reduction compared with the conventional orthogonal access (i.e., OFDMA). In this work, the effects of BAA on learning performance are quantified targeting a single-cell random network. First, we derive two tradeoffs between communication-and-learning metrics, which are useful for network planning and optimization. The power control (``truncated channel inversion'') required for BAA results in a tradeoff between the update-reliability [as measured by the receive \emph{signal-to-noise ratio} (SNR)]  and the expected update-truncation ratio. Consider the scheduling of cell-interior devices to constrain path loss. This gives rises to the other tradeoff between the receive SNR and fraction of data exploited in learning. Next, the latency-reduction ratio of the proposed BAA with respect to the traditional OFDMA scheme is proved to scale almost linearly with the device population. Experiments based on a neural network and a real dataset are conducted for corroborating the theoretical results. In addition, we discuss the extensions of BAA to acquire safety against adversary attacks and integrate beamforming for enhancing cell-edge links. 

\end{abstract}

\section{Introduction}

%\subsection{Overview of edge intelligence and edge learning}
The traffic in mobile Internet is growing at a breath-taking rate due to the extreme popularity of mobile devices (e.g., smartphones and sensors). Analysis shows that there will be 80 billions of devices connected to Internet by 2025, resulting in a tenfold traffic growth compared with 2016 \cite{Poggi2017datasource}. The availability of enormous mobile data and recent breakthroughs in \emph{artificial intelligence} (AI) motivate researchers to develop AI technologies at the network edge. Such technologies are collectively called \emph{edge AI} and drive the latest trend in machine learning, i.e., \emph{edge learning}, that concerns training of edge-AI models via computation at edge servers and devices \cite{zhu2018towards,Mao2017MECsurvey,wang2018edge}. The migration of learning from central clouds towards the edge allows edge servers to have fast access to real-time data generated by edge devices for fast training of AI models. In return, downloading the models from servers to devices in proximity provision the latter intelligence to respond to real-time events.  While computing speeds are growing rapidly, wireless transmission of high-dimensional data by many devices suffers from the scarcity of radio resources and hostility of wireless channels, resulting in a communication bottleneck for fast edge learning \cite{jordan2018communication,mcmahan2016communication}.
This calls for the design of low-latency multi-access schemes that integrate techniques from two different areas, namely distributed learning and wireless communication. 

% due to the emergence of powerful AI chips, e.g., \emph{graphics processing units} (GPUs) and \emph{tensor processing units} (TPUs), wireless data acquisition suffers from scarcity of radio resources and hostility of wireless channels, resulting in a communication bottleneck for fast edge learning \cite{jordan2018communication,mcmahan2016communication}.

%Research on edge learning is cross-disciplinary as it merges two areas: computer science and wireless communication. They concern two key aspects of edge learning, namely fast data processing and fast data acquisition from edge devices (smartphones and sensors), respectively. The two aspects cannot be decoupled as their performances are interwound under a common goal of fast learning. While computing speeds are growing rapidly, wireless data acquisition suffers from scarcity of radio resources and hostility of wireless channels, resulting in a bottleneck for fast edge learning \cite{mcmahan2016communication,jordan2018communication}.

\begin{figure}[tt]
\centering
\includegraphics[width=16cm]{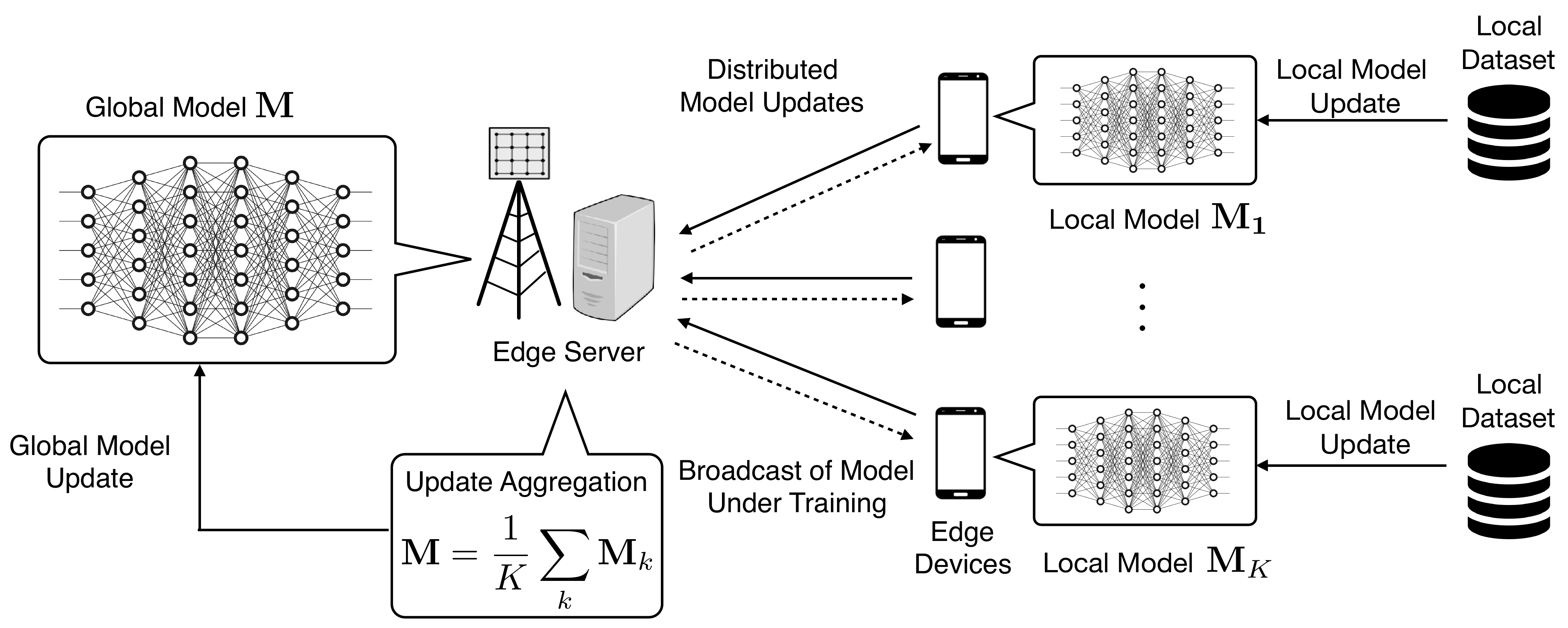}
\caption{Federated edge learning from wirelessly distributed data.}
\label{Fig:1}
\vspace{-3mm}
\end{figure}

In this work, we propose one such scheme, called \emph{broadband analog aggregation} (BAA), for low-latency implementation of a popular distributed-learning framework, called \emph{federated learning} \cite{mcmahan2016communication,konevcny2016federated}, in a wireless network, referred to as \emph{federated edge learning} (FEEL). As illustrated in Fig \ref{Fig:1}, a key operation of FEEL is to aggregate (or average) local models trained on devices to update the global model at a server. The BAA realizes the operation over a broadband multi-access channel by exploiting  simultaneous transmission and the resultant waveform superposition. This leads to dramatic latency reduction compared with the conventional orthogonal access. In this work, we develop the BAA framework by deriving the tradeoffs between a set of communication-and-learning metrics and quantifying the latency reduction compared with the conventional design.

\vspace{-3mm}
\subsection{Federated Edge Learning and Multi-Access}
\vspace{-1mm}
As mentioned, FEEL is a recently developed distributed-learning framework that preserves user-privacy by avoiding direct data uploading. To this end,  a typical federated-learning algorithm alternates between two phases, as shown in Fig. \ref{Fig:1}. One is to aggregate distributed updates over a multi-access channel and apply their average to update the AI-model at the edge server. The other is to broadcast the model under training to allow edge devices to compute its updates using local datasets and then transmit
the updates to the server for aggregation. The iteration continues until the global model converges and each iteration is called a \emph{communication round}. The updates computed locally at edge devices can be either the model parameters \cite{mcmahan2016communication} or gradient vectors \cite{konevcny2016federated}, giving rise to two implementation approaches, i.e., \emph{model-averaging} and \emph{gradient-averaging}.

%Model or gradient update uploading in federated learning can still be bandwidth-consuming as an AI model usually comprises millions to billions of parameters (so do the corresponding gradient vectors). Thereby, the local-updates by thousands of edge devices may easily congest the air-interface, making it a bottleneck for agile edge learning. The situation is further exacerbated by the fact that, to perform model averaging, the edge server needs to wait for the reception of all individual model updates. The synchronization requirement will incur unnecessary delay due to the idle time wasted on waiting for the straggling devices (the devices with slower processors or having more local data to process). 

%{\color{red} Para 2: shorten to emphasize the main stream approach in CS - reduce communication rounds.}

%{\color{red}
In view of high dimensionality in updates (each constitutes e.g., millions of parameters), a main theme in  the FEEL research is to develop communication-efficient strategies for fast update-uploading to accelerate learning.
There exist three main approaches. The first addresses the \emph{straggler issue}, namely that the slow devices (stragglers) dominate the overall latency due to update synchronization required for aggregation. To reduce latency, a partial averaging scheme is proposed in \cite{chen2016revisiting} where only a portion of updates from those fast-responding devices are used for global model updating, while those from stragglers are discarded. Later, the design was enhanced by coding the updates such that the full update-averaging can be still realized using only a portion of coded updates \cite{tandon2017gradient}. 
The second approach also aims at reducing the number of transmitting devices, but the scheduling criterion is update significance instead of computation speed \cite{kamp2018efficient,chen2018lag}.
If FEEL is implemented based on model averaging, the update significance is measured by the \emph{model variance} which indicates the divergence of a particular local model from the average across all local models \cite{kamp2018efficient}.
On the other hand, if the gradient averaging is employed, the update significance is measured by \emph{gradient divergence} that reflects the level of change on the current gradient update w.r.t. the previous one \cite{chen2018lag}. The last approach focuses on update compression by exploiting the  sparsity of  gradient updates \cite{aji2017sparse,lin2017deep}.
%}
% most of the gradient elements are insignificant and thus can be truncated without harming the model accuracy  

%{\color{red} Para 3: our approach is to apply elaborate and advanced communication techniques for reducing latency in each commmunication round. Describe the multi-access challenge (refer to RGC proposal). Then end by transit to AirComp.}

%{\color{red}
The prior work by computer scientists focuses on reducing the number of updating devices and compressing the information for transmission. It represents a computer-science approach for tackling the communication-latency problem in the FEEL systems. 
Wireless channels therein are abstracted as ``bit pipes'' that overlook the possibility of exploiting the channels' sophisticated properties (e.g., fading, multi-access and broadcasting, and spatial multiplexing) for latency reduction. Thus, a more direct and perhaps more fundamental approach for solving this communication problem is to develop wireless communication techniques to support low-latency FEEL. 
We adopt the new approach in this work and focus on designing a multi-access scheme for communication-efficient FEEL. The classic orthogonal-access schemes (e.g., OFDMA or TDMA) have been designed for supporting independent links. Their applications to edge learning can cause the multi-access latency to scale linearly with the number of edge devices and thus are inefficient. To overcome the drawback, we propose the low-latency BAA scheme for leveraging simultaneous broadband transmission to implement update aggregation ``over-the-air'' in the FEEL systems.

\subsection{Over-the-Air Computation} 
The current BAA scheme builds on the classic idea of over-the-air computation (AirComp). 
The idea of AirComp can be traced back to the pioneering work studying functional computation in sensor networks \cite{GastparTIT2007}. The design relies on structured codes (i.e., lattice codes) to cope with channel distortion introduced by the multi-access channel. The significance of the work lies in its counter-intuitive finding that ``interference" can be harnessed to help computing. It was subsequently discovered in \cite{GastparTIT2008} that simple analog transmission without coding but with channel pre-equalization can achieve the minimum distortion if the data sources are independent and identically distributed (i.i.d.) Gaussian.
Nevertheless, coding can be still useful for other settings if the sources follow more complex distributions such as bivariate Gaussian \cite{wagner2008rate} and correlated Gaussian \cite{SoundararajanTIT2012}. The satisfactory performance and simplicity of analog AirComp has led to an active area focusing on robustness design and performance analysis \cite{GoldsmithTSP2008,GoldenbaumWCL2014,C.H.WangTSP2011,GoldenbaumICC2015}. In particular, techniques for distributed power control and robust AirComp against channel estimation errors are  proposed in \cite{GoldsmithTSP2008} and \cite{GoldenbaumWCL2014}, respectively. Theoretical analysis on the AirComp outage performance under a distortion constraint and the computation rate, defined as the number of functional values computed per time slot, were provided in \cite{C.H.WangTSP2011} and \cite{GoldenbaumICC2015}, respectively.  Another vein of research focuses on transforming AirComp from theory into practice by prototyping \cite{KatabiAirComp2016} and addressing the practical issue of transmission synchronization over sensors \cite{GoldenbaumTCOM2013,KatabiAirshare2015}. In \cite{GoldenbaumTCOM2013}, the authors proposed to modulate the data into transmit power to relax the synchronization requirement. As a result, only coarse block-synchronization is required for realizing AirComp. An alternative scheme, called \emph{AirShare}, is developed in \cite{KatabiAirshare2015} which broadcasts a shared clock to all devices so as to enforce synchronization.

Advancing beyond scalar-valued function computation, the latest trend in the area also explores multiple-input-mutiple-output (MIMO) techniques to enable vector-valued function computation \cite{GuangxuIOT2018,XYTWC2018,DZTWC2018}, referred as MIMO AirComp. In particular, a comprehensive framework for MIMO AirComp that consists of beamforming optimization and a matching limited-feedback design is proposed in \cite{GuangxuIOT2018}. The framework was extended in subsequent work to wirelessly-powered AirComp system \cite{XYTWC2018}, where the beamformer was jointly optimized with the wireless power control to further reduce the AirComp distortion, and massive MIMO AirComp system \cite{DZTWC2018}, where a reduced-dimension two-tier beamformer design was developed by exploiting the clustered channel structure to reduce the channel-feedback overhead and signal processing complexity. It is also worth mentioning that, while AirComp is mostly deployed in computation-centric sensor networks as discussed above, the AirComp operation has been also leveraged in rate-maximization cellular systems such as two-way relaying \cite{GastparTIT2011} and MIMO lattice decoding \cite{SakzadTWC2013}.

%{\color{red}Revise last paragraph of Section 1.b to convey the following message: prior work focuses on narrowband channels since sensor networks typically require low-rate transmission and are allocated ``side links'' in practical systems e.g., LTE. For us, we consider broadband channels due to transmission of high dimensional updates in FEEL systems. Moreover, in our context, studying the effects of AirComp on edge learning is also new, which is irrelevant for prior work.}

%{\color{red}
Prior work on AirComp targets sensor networks and thus focuses on narrow-band systems only. The reason is that  sensor networks typically require low-rate transmission and communicate over a dedicated narrow-band in practical systems, e.g., LTE for narrow-band Internet-of-Things (NB-IoT) \cite{chen2017narrow}. In contrast, we consider broadband channels due to the transmission of high-dimensional updates in the FEEL systems. However, the solutions for broadband AirComp do not exist in the existing literature. Such design as well as the study of the effects of AirComp on the edge-learning performance is an area largely uncharted. This motivates the development of the BAA scheme and its performance analysis in the current work.
%without being explored in the prior work that investigates only the basic AirComp metrices (e.g., computation error and throughput). 
%}

% The abundant spectral degrees-of-freedom (DoF) provided by the broadband system can be also exploited for high-dimensional vector-valued function computation, e.g., update-aggregation in FEEL. Nevertheless, to the best of the authors knowledge, broadband AirComp has not been reported in the existing literature. The underlying reason is that the conventional broadband solution, i.e., the orthogonal frequency division multiplexing (OFDM) architecture, and the power control policy are designed for \emph{digital} communication aiming at \emph{rate maximization}, which cannot be directly applied for \emph{analog} AirComp with a distinct objective of \emph{computation-error minimization}. Therefore a new broadband AirComp solution is in demand.  

%XXX: The performance of the broadband AirComp was largely unknown.
%
%XXX: The conventional OFDM architecture not applicable?
%
%XXX: Modify the following to show the prior work  focus on only narrow band but not broad band has been considered. 
%
%XXX:  Also check Chen Li's paper on broadband AirComp to see what is the challenging point there. 
%
%XXX: For gradient averaging, distributed setting is equivalent to centralized setting so non-uniform data partition and uniform one are the same. 

%\vspace{-6mm}
\subsection{Contribution and Organization}
We consider the implementation of FEEL in a single-cell wireless network. The BAA scheme is proposed to reduce the communication latency in the network, which is described as follows. Each device transmits a high-dimensional update (local model)
in blocks over a broadband channel, using linear analog modulation for modulating individual parameters and \emph{orthogonal frequency division multiplexing} (OFDM) for partitioning the channel into sub-channels. Realizing over-the-air update aggregation requires the received model parameters from different devices to have identical amplitudes, called \emph{amplitude alignment}. This is achieved by broadband channel inversion at each device. The channel inversion and simultaneous analog OFDM transmission by a set of scheduled devices allow the server to receive the desired average of the local models/updates computed at the devices. 

Based on the scheme, we develop the BAA framework using a random network model where a number of edge devices are randomly distributed in a disk area. To describe our findings, it is necessary to introduce some metrics related to the FEEL-network performance as follows:

\begin{enumerate}
\item \underline{\emph{Receive SNR}}: Given amplitude alignment, the receive \emph{signal-to-noise ratios} SNRs for updates transmitted by different devices are identical. The metric is one quality measure of model update in FEEL. 

\item \underline{\emph{Truncation ratio}}: This is another update-quality measure. It refers to the expected ratio of model parameters being truncated due to channel inversion at a device under a transmit-power constraint. 

\item \underline{\emph{Fraction of exploited data}}: It refers to the fraction of the distributed dataset exploited in learning. Given uniform data distribution over devices, the metric is equal to the fraction of scheduled devices. 
\end{enumerate}

The findings and the contributions from the framework development are summarized as follows:

\begin{itemize}
\item {\bf Two Communication-and-Learning Tradeoffs}: The first tradeoff as derived in closed-form is between the receive SNR and the truncation ratio, called the \emph{SNR-truncation tradeoff}. The tradeoff also shows that the receive SNR is limited by the path-loss of the furthest device from the server, which is due to the said amplitude alignment among devices. This suggests that the receive SNR can be improved via scheduling cell-interior devices for FEEL. However, this causes data loss and thereby gives rise to the second tradeoff, namely the one between the receive SNR and the fraction of exploited data. It is referred to as \emph{(update)-reliability-(data)-quantity tradeoff}. The above two tradeoffs are fundamental for FEEL systems with BAA and can be useful tools for further research in this direction.  In addition, to improve this tradeoff by copying with data deficiency, we propose two scheduling schemes: one is to exploit mobility and the other to alternate the scheduling of cell-interior and cell-edge devices. 

\item{\bf Communication-Latency Analysis}: The latency reduction of BAA is quantified with respect to (w.r.t.) the conventional multi-access scheme, namely \emph{orthogonal frequency division multiple access} (OFDMA) with digital modulation. 
 The latency-reduction ratio is proven to increase with $K$, the number of scheduled devices, as $O\l(\frac{K}{\log K} \r)$. The result shows that BAA is a promising solution for low-latency FEEL with many devices. 

\item{\bf Experiments}: The FEEL system is implemented in software for an AI application of handwritten-digit recognition, where the AI-model is based on a neural network and a real image dataset. Experimental results demonstrate the derived communication-learning tradeoffs. Moreover, the results confirm the dramatic communication-latency reduction achieved by BAA w.r.t. the conventional design. 

\item {\bf Extensions}: Extensions of the BAA scheme are also presented to address two issues, namely security against adversarial attacks and beamforming for improving link reliability of cell-edge devices. 
\end{itemize}

Last, it is worth mentioning that upon the completion of this work, interesting parallel work \cite{Deniz2019analogSGD,yang2018federated} was also reported that share the common theme of applying AirComp to update aggregation in FEEL. Nevertheless,  the specific systems and designs therein differ from the current work. Both paralelled work targets narrow-band single/multi-antenna channels and focuses on algorithmic design to improve system performance. Specifically, a source-coding algorithm that exploits the sparsity in gradient update is proposed in \cite{Deniz2019analogSGD} for compressed update transmission. A device-selection algorithm is developed in \cite{yang2018federated} to maximize the number of scheduled devices under a update-distortion constraint.

\emph{Organization}: The remainder of the paper is organized as follows. Section II introduces the system and channel models. Section III presents the proposed BAA scheme and motivates the user scheduling problem. Practical scheduling schemes for BAA is presented in Section IV and the involved tradeoff is quantified. The latency performance of the proposed BAA is analytically compared with the digital counterpart in Section V. Section VI shows the experimental results using real dataset.  Discussion on possible extensions is provided in Section VII,  followed by concluding remarks in Section VIII.

\section{System Model}
\subsection{Federated Edge Learning System}
We consider a FEEL system comprising a single edge server and $K$ edge devices as  shown in Fig.~\ref{Fig:1}. A shared AI model (e.g., a classifier), represented by the parameter set $\bw$, is trained collaboratively across the edge devices. Each device collects a fraction of labelled training data via the interaction with its own user, constituting a local dataset. 
 
To facilitate the learning, the loss function measuring the model error is defined as follows.
 Let ${\cal D}_k$ denote the local dataset collected at the $k$-th edge device. The \emph{local loss function} of the model vector $\bw$ on ${\cal D}_k$ is given by
\begin{align}\label{eq:local_loss} (\text{Local loss function}) \qquad
F_k(\bw) = \frac{1}{|{\cal D}_k|} \sum_{(\bx_j, y_j) \in {\cal D}_k} f(\bw, \bx_j, y_j),
\end{align}
where $f(\bw, \bx_j, y_j)$ is the sample-wise loss function quantifying the prediction error of the model $\bw$ on the training sample $\bx_j$ w.r.t. its ground-true label $y_j$. 
For convenience, we rewrite $f(\bw, \bx_j, y_j)$ as $f_j(\bw)$ and  assume uniform sizes for local datasets: $|{\cal D}_k| \equiv D$, for all $k$. Then, the \emph{global loss function} on all the distributed datasets can be written as
\begin{align} (\text{Global loss function}) \qquad
F(\bw) = \frac{\sum_{j \in \cup_k {\cal D}_k }f_j(\bw)}{|\cup_k {\cal D}_k|} = \frac{1}{K} \sum_{k = 1}^K F_k(\bw).
\end{align}

The learning process is thus to minimize the global loss function $F(\bw)$, namely,
\begin{align}\label{eq:learning_prob}
\bw^* = \arg \min F(\bw). 
\end{align}
One way for computing $F(\bw)$ is to directly upload all local data, which causes the privacy issue.  
To tackle the issue, the FEEL framework is employed to solve the problem in \eqref{eq:learning_prob} in a distributed manner. We focus on the model-averaging implementation in the subsequent exposition while the same principle also applies to the alternative implementation based on gradient-averaging \cite{konevcny2016federated}.

For implementing FEEL, in each communication round, say the $n$-th round, the edge server broadcasts the current model under training $\bw[n]$ to all edge devices. Starting from $\bw[n]$, each device updates its own model by running $\tau$-step ($\tau \geq 1$) \emph{stochastic gradient descent} (SGD) towards minimizing the loss function defined in \eqref{eq:local_loss}. Mathematically, for device $k$, a single-step SGD updates the local model $\bw_k$ based on the following equation:  
 \begin{align}\label{eq:local_update}(\text{Local model updating}) \qquad
\bw_k[n+1] = \bw[n] - \eta \nabla F_k(\bw[n]),
\end{align}
where $\eta$ is the step size and $\nabla$ represents the gradient operator. Then, a $\tau$-step SGD repeats the updating rule in \eqref{eq:local_update} for $\tau$ times. Upon its completion,  the local model-updates are sent to the edge server for averaging and updating the global model $\bw$ as follows:
 \begin{align}\label{eq:model_averaging} (\text{Global model updating}) \qquad
\bw[n + 1] = \frac{1}{K} \sum_{k=1}^K \bw_k[n + 1].
\end{align} 
The learning process involves the iteration between \eqref{eq:local_update} and \eqref{eq:model_averaging} until the model converges. 

As observed from \eqref{eq:model_averaging}, it is only the aggregated model, i.e., 
\begin{align}\label{eq:model_aggregation} (\text{Model aggregation}) \qquad
\by = \sum_{k=1}^K \bw_k,
\end{align} 
instead of individual model-updates $\{\bw_k\}$, needed at the edge server for model averaging. This motivates the low-latency BAA scheme exploiting AirComp 
%for ``one shot'' model aggregation by allowing simultaneous uploading elements of $\{\bw_k\}$ 
as presented in Section \ref{sec:analog_model_aggregation}.

\subsection{Broadband Channel and Update Transmission} 
The uploading of model-updates from edge devices to the server is through a broadband multi-access channel. To cope with the frequency selective fading and inter-symbol interference, the OFDM modulation is adopted to divide the whole bandwidth $B$ to $M$ orthogonal sub-channels.  To exploit AirComp for low-latency model aggregation, model-updates are amplitude-modulated for analog transmission. Also, each sub-channel is dedicated for one model-parameter transmission. 

During the model updating phase, all devices transmit simultaneously over the whole available bandwidth. We assume symbol-level synchronization among the transmitted devices through a synchornization channel (e.g., ``timing advance'' in LTE systems \cite{TALTE}). 
 Let $\bw_k = [w_{k,1}, w_{k,2}, \cdots, w_{k,q}]^T$ denote the $q\times 1$ local model-update from the $k$-th device, where  $q$ also denotes the number of model parameters. At each communication round, the model-updating duration consists of $N_s = \frac{q}{M}$ OFDM symbols. In particular, the $i$-th aggregated model parameter, denoted by $y_i$, with $i = (t-1)M + m$, received in the $m$-th sub-carrier at the $t$-th OFDM symbol, is given by
 \begin{align}\label{channel_model}
 y_i = \sum_{k=1}^K r^{-\frac{\alpha}{2}}_k h^{(m)}_k[t] p^{(m)}_k[t] w_{k,i} + z^{(m)}[t], \qquad \forall i
 \end{align}
 where $r^{-\frac{\alpha}{2}}_k$ captures the path-loss of the link between device $k$ and the edge server, with $r_k$ denoting the distance between them and $\alpha$ representing the path-loss exponent; the small-scale fading of the channel is captured by $h^{(m)}_k[t]$ which follows Rayleigh fading and is \emph{identically and independently distributed} (i.i.d.) over the indexes of $k,m,t$, yielding  $h^{(m)}_k[t] \sim {\cal CN}(0,1).$\footnote{We assume a highly frequency selective channel for ease of exposition. For a frequency correlated channel, it can be modelled by frequency block fading where sub-channels within a same block have identical gains and the gains over different blocks are i.i.d. \cite{adireddy2002optimal}. The main results hold if the definition of (parameter) truncation ratio is modified as the ratio of truncated frequency blocks.} $\{p^{(m)}_k[t]\}$ are the associated power control policies on the transmitted updates to be designed in the sequel. Last,  $z^{(m)}[t]$ models the i.i.d. \emph{additive white Gaussian noise} (AWGN) following ${\cal CN}(0,1)$. 
 For ease of notation, we skip the OFDM symbol index $t$ in the subsequent exposition whenever no confusion is incurred.
To facilitate the power-control design and reduce transmission power, the symbols are normalized to have zero mean and unit variance, i.e., ${\sf E}(\bw_k \bw_k^H) = \bI$, where the normalization factor for each model dimension is uniform for all devices and can be inverted at the edge server.

%To implement the AirComp-based model aggregation, the power adaptation factor $p_k$ needs to be designed according to the channel coefficient $h_k$ to ensure the desired aggregation of the model-updates at the edge server.

The power allocation over sub-channels, $\{p_k^{(m)}\}$, will be adapted to the channel coefficients, $\{h_k^{(m)}\}$, for implementing  the BAA as presented in the sequel. The transmission of each device is subject to the long-term transmission power constraint:
\begin{align}\label{power_constraint1}
{\sf E} \l[{ \sum_{m=1}^M |p_k^{(m)}(h_k^{(m)}) } |^2  \r] \leq P_0, \qquad \forall k,
\end{align}
where the the expectation is taken over the random channel coefficients. 
Since channel coefficients are i.i.d. over different sub-channels, the above power constraint reduces to
\begin{align}\label{power_constraint2}  (\text{Power constraint}) \qquad
{\sf E} \l[  | p_k^{(m)}(h_k^{(m)})|^2 \r]  &\leq \frac{P_0}{M}, \qquad \forall k.
\end{align} 

%A power control policy targeting AirComp of \eqref{eq:model_aggregation} is developed under the above constraint in the sequel.

\begin{figure}[tt]
\centering
\includegraphics[width=10cm]{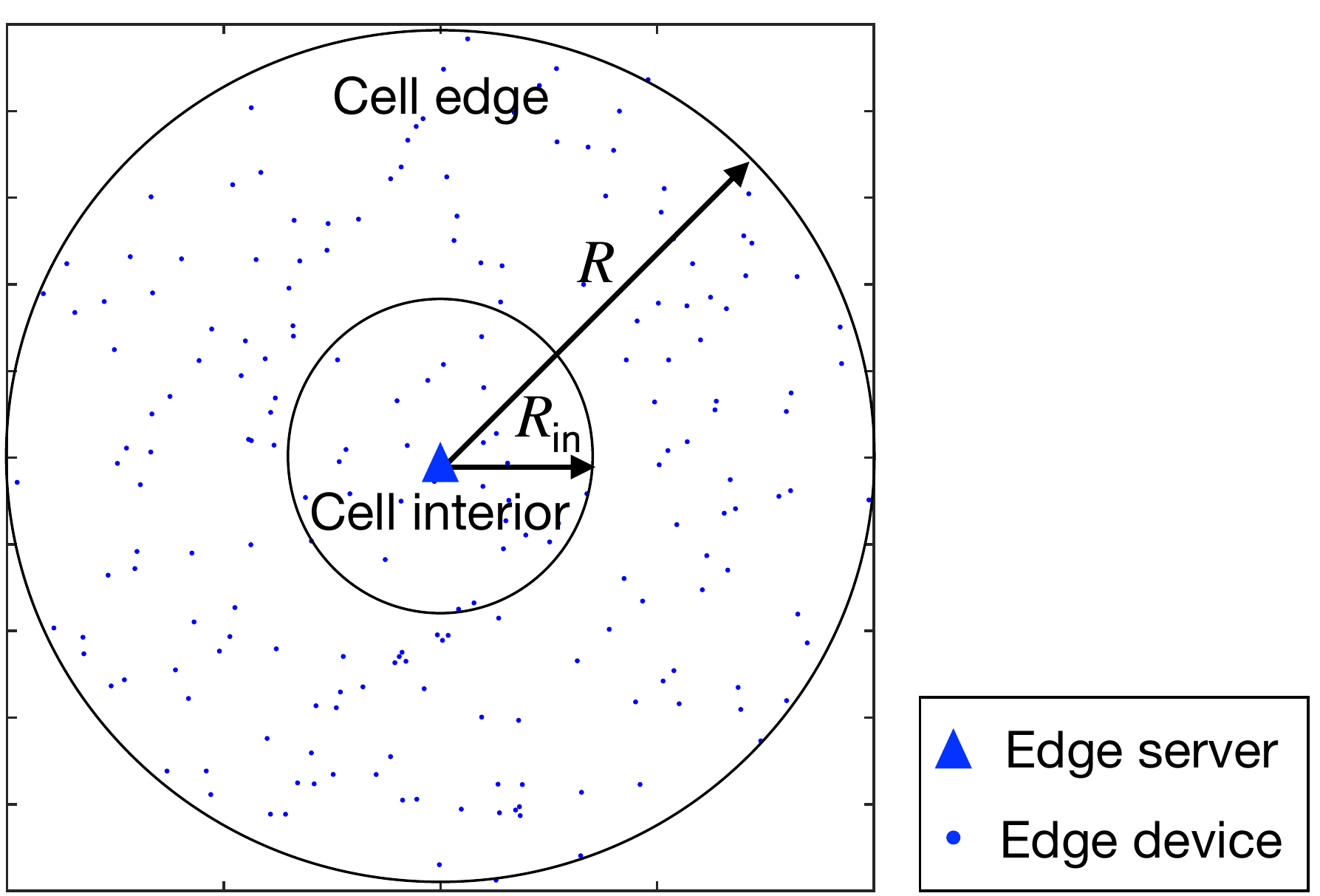}
\caption{A single-cell edge learning network with uniformly-distributed edge devices.}
\label{Fig:uniform_dist}
\end{figure}
 
\subsection{Network Topology} \label{sec:uniform_user_dist}
We consider a single-cell network distribution in a disk. Specifically, the edge devices are i.i.d. distributed over a disk centred at the edge server with
a cell-radius $R$. Thus the \emph{probability density function} (PDF) of the distance $r_k$ is given by
\begin{align}\label{pdf_r_k}
f_{r_k}(r) = \frac{2r}{R^2}, \qquad 0 \leq r \leq R.
\end{align}

\noindent Fig. \ref{Fig:uniform_dist} illustrates one realization of the random network. The cell is divided into two non-overlapping parts: cell-interior and cell-edge. Specifically, the area within a range of distance $R_{\sf in}$ from the server is referred to as the  \emph{cell-interior} while the area outside the range as the \emph{cell-edge}.

%\begin{align}
%f_{r_k}(r) = \frac{2r}{R^2 - r_0^2}, \qquad r_0 \leq r \leq R,
%\end{align}
%where $r_0$ denotes the guard interval that specifies the nearest distance between the devices and edge server. 

%Analog amplitude modulation is applied to the model updates to facilitate the low-latency model aggregation by AirComp as elaborated in the sequel. 

% As specified in the sequel, the implementation of AirComp-based model aggregation requires the model updates to be amplitude-modulated. 

%denoted as $\{(\bx_k, c_k)\}$ with $\bx_k \in \mathbb{R}^p$ representing the $k$-th data sample,  $p$ its dimensions, and $c_k\in \{1, 2, \cdots, C\}$ its label. To this end, edge devices share the wireless channel in a time division manner and take turn to transmit local data to the server. Note that a label has a much smaller size than a data sample (e.g., a $0-9$ integer versus a vector of a million real coefficients). Thus two separate channels are planned: a low-rate   \emph{label channel} and a high-rate  \emph{data channel}. The former is assumed  noiseless for simplicity. Reliable uploading of data samples over the noisy and fading channel is  the bottleneck of  wireless data acquisition and  the focus of  this work.  Time is slotted into symbol durations, called \emph{slots}. Transmission of a data sample requires $p$ slots, called a \emph{symbol block}.

\section{Broadband Analog Aggregation:  Scheme and Properties}\label{sec:analog_model_aggregation}
In this section, the proposed BAA scheme for FEEL is first presented. Then the resultant SNR-truncation tradeoff as mentioned is derived via analyzing the receive SNR and update-truncation ratio.

%\begin{figure}[tt]
%\centering
%\includegraphics[width=16cm]{System_diagram.pdf}
%\caption{Transmitter design at edge devices for broadband model aggregation.}
%\label{Fig:system_diagram}
%\end{figure}

\subsection{The Scheme of Broadband Analog Aggregation}
%Due to the said superposition property, simultaneously transmitted analog-waves by different devices will be automatically superposed at the receiver but weighted by the channel coefficients. Therefore,  to enforce the desired model aggregation over the air, AirComp essentially involves the application of \emph{analog amplitude modulation} to model coefficients and a power control for \emph{pre-channel-compensation} at the transmitter.

% As mentioned earlier, the key to implement AirComp is to properly control the transmit power of each device such that the model-updates can be aggregated coherently at the edge server without being weighted by the channel coefficients. 
 
\subsubsection{Transmitter Design}
\begin{figure}[tt]
 \centering
   \subfigure[Transmitter design for edge devices]{\label{subfig:tx}\includegraphics[width=0.95\textwidth]{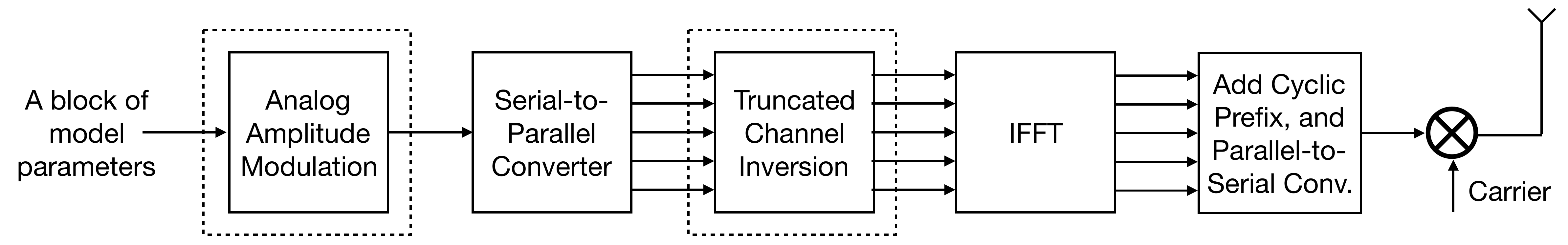}}
  \hspace{0.25in}
  \subfigure[Receiver design for edge server]{\label{subfig:rx}\includegraphics[width=0.85\textwidth]{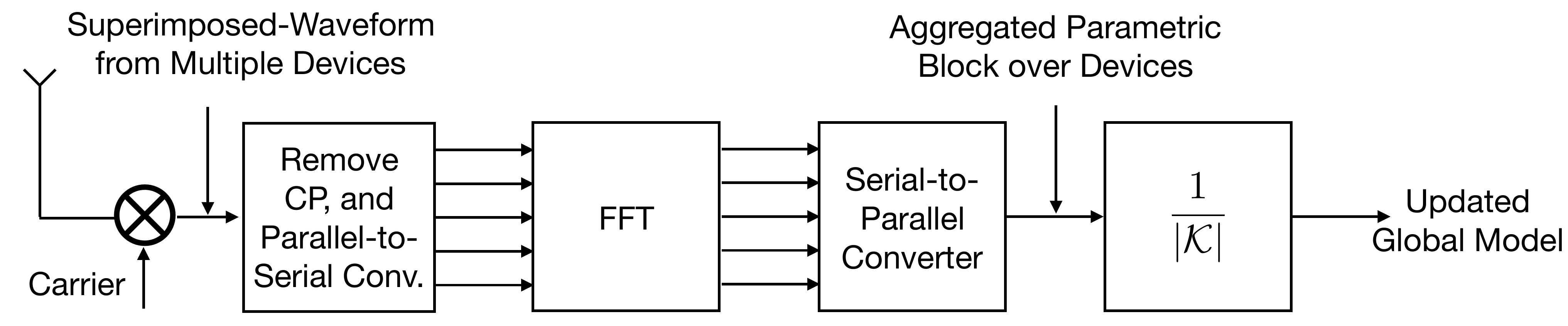}}
   \vspace{3mm}
  \caption{Transceiver design for broadband model aggregation.}
  \label{Fig:Fig:system_diagram}
\end{figure}

To enable BAA, the transmitter design for edge devices is illustrated in  Fig. \ref{subfig:tx}.  Essentially, the design differs from the classic  OFDM transmitter by replacing digital modulation (e.g., QAM) with linear analog modulation and adding channel-inversion power control, as highlighted in  Fig. \ref{subfig:tx}. 
%in Fig. \ref{subfig:tx}.

The new signal-processing operations in the transmitter are described as follows. The local-model parameters are first amplitude-modulated into symbols. The long symbol sequence is divided into blocks. Each is transmitted in a single OFDM symbol with one parameter over one frequency sub-channel. Assuming perfect CSI at the transmitter, sub-channels are inverted by power control so that model parameters transmitted by different devices are received with identical amplitudes, achieving amplitude alignment at the receiver as required for BAA.  Nevertheless, a brute-force approach is inefficient if not impossible under a power constraint since some sub-channels are likely to encounter deep fades. To overcome the issue, we adopt the more practical \emph{truncated channel inversion}. To be specific, a sub-channel is inverted only if its gain exceeds a so called \emph{power-cutoff threshold}, denoted by $g_{\sf th}$, or otherwise allocated zero power. Then the transmission power of the $k$-th device on the $m$-th sub-channel, denoted as $p_k^{(m)}$, can be written as
\begin{align}\label{truncated_channel_inv} (\text{Truncated channel inversion}) \qquad
 p_k^{(m)} = \l\{\begin{aligned} 
& \frac{\sqrt{\rho_k}}{r_k^{-\frac{\alpha}{2}} h_k^{(m)}}, && |h_k^{(m)}|^2 \geq g_{\sf th} \\
&0, &&  |h_k^{(m)}|^2 < g_{\sf th},
\end{aligned}
\r.
 \end{align}
where $\rho_k$ is a scaling factor set for ensuring the average-transmit-power constraint in \eqref{power_constraint2}. One can see from \eqref{channel_model} that, $\rho_k$ also determines the  receive SNR of the model-update from each device.

We remark that the policy can cause the loss of those model parameters that are mapped to the truncated sub-channels. To measure the loss, we define the truncation ratio as  $\zeta = \frac{\text{ \# of truncated parameters}}{\text{\# of total model-update parameters}}$  and analyze it in the sequel. Other operations in Fig. \ref{subfig:tx} follow the conventional design. Their details are omitted for brevity. 

\subsubsection{Receiver Design}
Fig. \ref{subfig:rx} shows the receiver design for edge server. It has the same architecture as the conventional OFDM receiver
except that the digital demodulator is replaced with a post-processing operator that scales the received signal to obtained the desired average model. However, the received signal is different between the current receiver and the conventional design as described below.

Consider an arbitrary communication round and a set of devices scheduled by the server to transmit their local models, which are represented by the index set ${\cal K}$. Given their simultaneous transmission, the server receives superimposed waveforms. By substituting the truncated-channel-inversion policy in \eqref{truncated_channel_inv} into \eqref{channel_model}, the server obtains the aggregated local-model block, denoted by a $M\times 1$ vector $\by[t]$,  at the serial-to-parallel converter output [see Fig. \ref{subfig:rx}] as: 
\begin{align}\label{channel_model2}
\by[t] = \sum_{k \in {\cal K}} \sqrt{\rho_k} \tilde\bw_{k}[t] + \bz[t], 
\end{align}
where $t$ is the index of local-model block (OFDM symbol)  as defined in \eqref{channel_model}. $\tilde\bw_{k}[t]$ is a truncated version of $\bw_{k}[t] = [w_{k, (t-1)M + 1}, \cdots, w_{k,tM}]^T$ with the truncated elements determined by the channel realizations according to \eqref{truncated_channel_inv} and represented by zeros. Note from \eqref{channel_model2} that $\{\sqrt{\rho_k}\}$ should be aligned to enforce the said amplitude alignment required for aggregation.  
%\begin{align}\label{channel_model2}
%y_i = \sum_{k \in {\cal K}} \sqrt{\rho_k} w_{k,i} + z, \qquad \forall i,
%\end{align}
Next, cascading all the $N_s$ blocks and scale the result by the factor $\frac{1}{|\cal K|}$ gives the desired updated global model. Then, the server initiates the next communication round by broadcasting the model to all devices or complete the learning process if the model converges.

\subsection{SNR-Truncation Tradeoff}
Targeting the BAA scheme, we show in this sub-section that there exists a tradeoff between the receive SNR (identical for all devices) and the truncation ratio defined in the preceding sub-section, which is regulated by the power-cutoff threshold in \eqref{truncated_channel_inv}.

%As seen in \eqref{truncated_channel_inv}, the power-cutoff threshold $g_{\sf th}$ is the key  parameter to be determined in the proposed BAA scheme with truncated channel inversion. Particularly, for each model-update, $g_{\sf th}$ determines a tradeoff between the \emph{receive SNR} and the \emph{truncation ratio},  defined by $\zeta = \frac{\text{ \# of truncated coefficients}}{\text{\# of total model-update coefficients}}$, as elaborated in the following.

First, by substituting \eqref{truncated_channel_inv} into \eqref{power_constraint2}, the maximum receive SNR of a model-update can be derived below.
\begin{lemma}[Maximum receive SNR]\label{prop:1} \emph{Consider the $k$-th edge device with the propagation distance $r_k$, the maximum receive SNR of the update transmitted by the device is bounded as}
\begin{align}\label{receive_power_k}
\rho_k \leq \frac{P_0}{M r_k^\alpha {\sf Ei}(g_{\sf th})}, 
\end{align}
\emph{where  ${\sf Ei}(x)$ is the exponential integral function defined as  ${\sf Ei}(x) = \int_x^\infty \frac{1}{t} \exp(-t) dt$. The equality is achieved when the device transmits with the maximum average power $P_0$.
}
\end{lemma}
\proof
Let $g_k = |h_k|^2$ denote the channel gain of the $k$-th link. Since the channel coefficient is Rayleigh distributed $h_k \sim {\cal CN}(0,1)$, it yields that $g_k$ follows the exponential distribution with unit mean. Then substituting \eqref{truncated_channel_inv} into \eqref{power_constraint2} gives 
\begin{align}
\frac{\rho_k}{r_k^{-\alpha}} \int_{g_{\sf th}}^\infty \frac{1}{g} \exp(-g) dg \leq \frac{P_0}{M}.
\end{align} 
 The desired result follows by invoking the definition of the exponential integral function. \endproof

Lemma \ref{prop:1} indicates that the maximum receive SNR for a model-update is limited by the propagation distance. 
For BAA implementation, near devices need lower their transmission power to achieve amplitude alignment with far devices. 
This results in a receive SNR identical for all devices, denoted as $\rho_0$. It follows from Lemma  \ref{prop:1} that
\begin{align}\label{aligned_power}
(\text{Receive SNR}) \quad  \rho_0 = \frac{P_0}{M r_{\max}^\alpha {\sf Ei}(g_{\sf th})}, 
\end{align}
where $r_{\max} = \max_k \{r_1, r_2, \cdots, r_K\}$ denotes the distance from the edge server to the furthest active device. 
The result suggests the need of limiting the distances of devices by scheduling, which is explored in Section \ref{sec:opportunistic scheduling}.

%Lemma \ref{prop:1} also specifies the monotone relation between the power-cutoff threshold $g_{\sf th}$ and the receive SNR for the uploaded update. 
Besides affecting the receive SNR, the power-cutoff threshold $g_{\sf th}$ also regulates the truncation ratio. In particular, when an update contains sufficiently many parameters, by law of large number, its truncation ratio is equal to the corresponding channel cutoff probability as derived below.

\begin{lemma}[Truncation ratio]\label{prop:2} \emph{When the model-update dimension $q \to \infty$,  the truncation ratio $\zeta$ is equal to the channel-cutoff probability:}
\begin{align}\label{eq:truncation_ratio}
\zeta \to {\sf Pr}(|h_k|^2 < g_{\sf th}) = 1-\exp(-g_{\sf th}), \qquad q \to \infty
\end{align}
\proof The result immediately follows from the exponential distribution of the channel gain.
\end{lemma}

Combining Lemmas \ref{prop:1} and \ref{prop:2}, we derive the said SNR-truncation tradeoff as follows.

\begin{proposition}[SNR-truncation tradeoff]\label{prop:snr_trunc_trade}
\emph{Given the BAA scheme, the relationship between the receive SNR $\rho_0$ and the truncation ratio $\zeta$ is specified by the following equation:}
\begin{align}
 \rho_0 = \frac{P_0}{M r_{\max}^\alpha {\sf Ei}(-\ln(1 - \zeta) )}, \qquad q \gg 1.
\end{align}
\end{proposition}

\begin{figure}[tt]
\centering
\includegraphics[width=10cm]{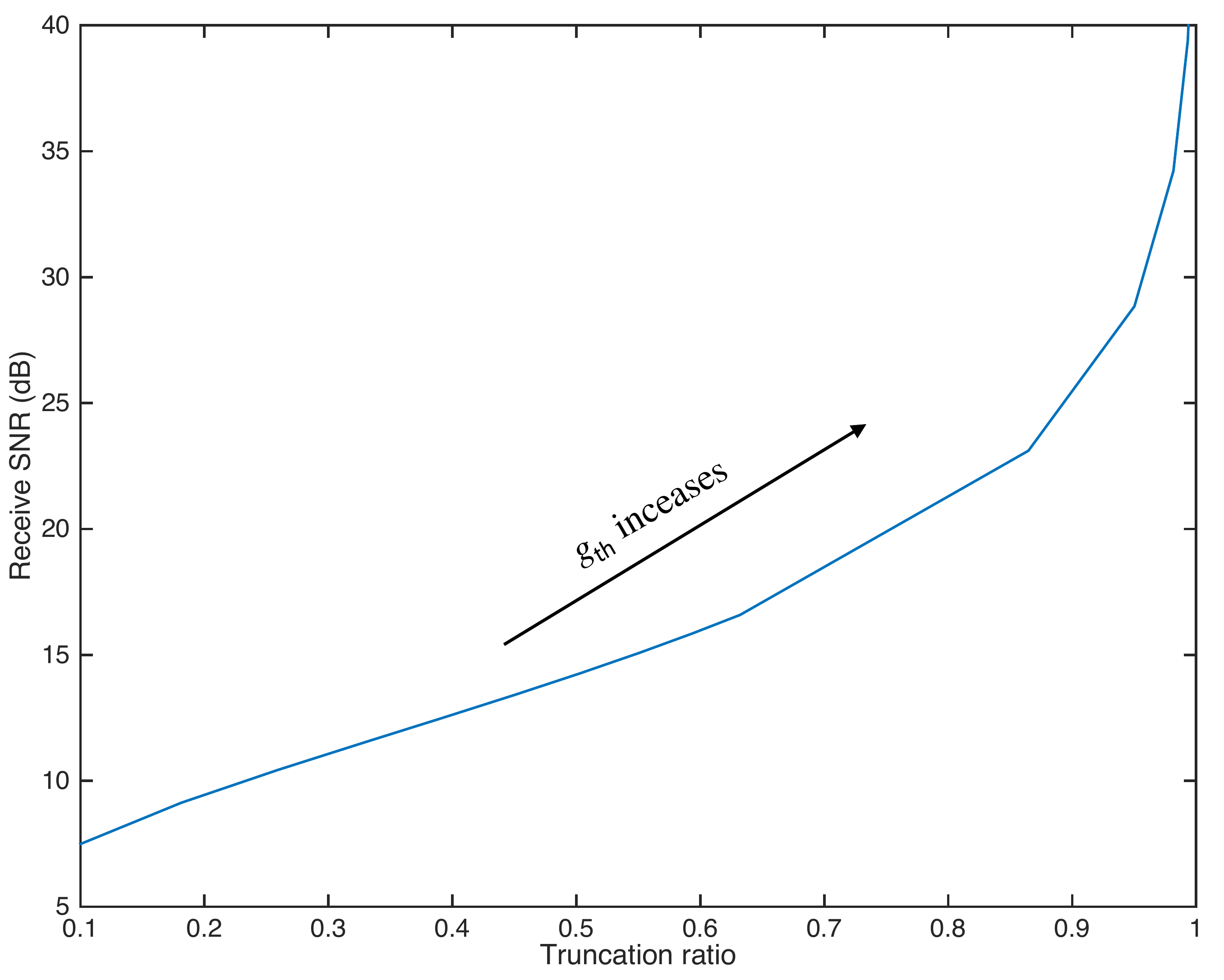}
\caption{Illustration of the SNR-truncation tradeoff, where we set $P_0 = 0.1$(W), $M =1000$, $r_{\max} = 100$, $N_0 = -80$ {dBm} and $\alpha = 3$.}
\label{Fig:2}
\vspace{-3mm}
\end{figure}
 
\begin{Remark}[SNR-truncation tradeoff and power-cutoff threshold] \emph{ 
%Proposition \ref{prop:2} quantifies how much SNR $\rho_k$ can be traded for if a truncation ratio of $\zeta$ is tolerated.
  An exemplary SNR-truncation tradeoff curve for typical system settings is plotted in Fig. \ref{Fig:2}. The said tradeoff is controlled by the power-cutoff threshold $g_{\sf th}$. Particularly, an increasing $g_{\sf th}$ tends to increase the receive SNR at the cost of more truncated parameters in model updates and vice versa. Since both affect the receive-update quality at the server, the threshold being too high or too low degrades the learning performance. Thus it is necessary to optimize the threshold, which is a design issue warranting further investigation. In experiments, the power-cutoff threshold is optimized numerically to optimize the learning performance by a grid search.
}
\end{Remark}

%Therefore, the selection of $g_{\sf th}$ is crucial to the ultimate learning performance. 
%In the subsequent learning performance evaluation, the power-cutoff threshold will be optimized numerically to maximize the learning performance by grid search.
%To provide an overview of the effect of  power-cutoff threshold on the aggregated model-update, a signal model illustrating the process of the BAA is shown in Fig. \ref{Fig:analog_model_aggregation}. 

%XXX: Mention the near-far problem in the AirComp, the key issue is the received signal power level is determined by the furthest users suffering from the strongest path-loss.
%
%XXX: Define the notion of update-SNR.

\section{Broadband Analog Aggregation: Scheduling}\label{sec:opportunistic scheduling}

%\begin{Remark}[User scheduling problem in AirComp]\label{Remark:user_scheduling} \emph{As mentioned, the receive SNR of the aggregated update is limited by the furthest user in the cell (see \eqref{aligned_power}), simply scheduling all the available users, referred as \emph{all-inclusive scheduling}, may compromise the reliability of the update substantially. 
%a large distance ratio and thus a highly inefficient use of available power at the nearby devices. On the other hand, scheduling only the nearby users may also compromise the learning performance due to the unexploitation of the data distributed at the cell-edge. 
%Therefore, it raises a tradeoff concerning inter-device coordination in the user-scheduling problem for broadband analog aggregation, that is to \emph{balance the resultant receive SNR of the aggregated update and the amount of data exploited in the update computation. }
%%may lead to a significant SNR penalty on the aggregated model-update, and thus deteriorate the learning performance. 
%}
%\end{Remark}

The preceding result in Proposition \ref{prop:2} shows that the bottleneck of the receive SNR of model updates is the device with the longest propagation distance. Then to ensure the update reliability, it is desirable to constrain the distance of active devices from the server. This motivates the following scheduling scheme.  
\begin{framed}
\vspace{-10pt} 
\begin{scheme}[Cell-interior scheduling]\label{scheme:1}
\emph{The edge server schedules only the cell-interior edge devices whose distances are no larger than a distance threshold $R_{\sf in}$.
}
\end{scheme}
\vspace{-10pt} 
\end{framed}

For the purpose of comparison, consider the baseline scheme of simply scheduling all available devices, called \emph{all-inclusive scheduling}. Compared with the baseline scheme, even though cell-interior scheduling improves the update reliability, it has the drawback of \emph{data deficiency} since the resultant model-training fails to exploit data at cell-edge devices. As a basic property of cell-interior scheduling, we derive in the sequel a tradeoff between the SNR gain (w.r.t. all-inclusive scheduling) and the fraction of exploited data, called the (update)-reliability-(data)-quantity tradeoff, by analyzing the two metrics. In the last subsection, schemes for coping with data deficiency are discussed.

\subsection{Fraction of Exploited Data.}
Assuming equal data partitioning among all edge devices, the fraction of exploited data is equal to the fraction of scheduled users,  which is derived as follows. 
Let $K_{\sf in}$ denote the number of scheduled devices within the range of $R_{\sf in}$.  The number $K_{\sf in}$ is a random variable whose distribution is parameterized by $R_{\sf in}$, $R$, and $K$ as derived below.
\begin{lemma}[Distribution of the number of scheduled devices]\label{lemma:1}
\emph{In cell-interior scheduling, given the distance threshold $R_{\sf in}$, the number of scheduled users follows a Binomial distribution with the \emph{probability mass function} (PMF) given by:
\begin{align}
{\sf Pr}(K_{\sf in} = k) = \binom{K}{k} \l(\frac{R_{\sf in}^2}{R^2} \r)^k \l( 1-\frac{R_{\sf in}^2}{R^2} \r)^{K-k},
\end{align}
}\proof
See Appendix \ref{app:lemma:1}. 
\endproof
\end{lemma}

Based on Lemma \ref{lemma:1}, one can easily obtain the expected fraction of exploited data as follows:
\begin{proposition}[Expected fraction of exploited data]\label{prop:F_DAT} \emph{Given cell-interior scheduling with the distance threshold $R_{\sf in}$, the expected fraction of exploited-data (or equivalently the fraction of scheduled-devices) is  given by 
}
\begin{align}
F_{\sf DAT} = {\sf E}\l(\frac{K_{\sf in}}{K} \r) = \l( \frac{R_{\sf in}}{R} \r)^2.
\end{align}
\end{proposition}

\vspace{-4mm}
\subsection{Receive SNR Gain}
To characterize the receive SNR gain, the expected receive SNRs for all-inclusive scheduling and cell-interior scheduling are analyzed. Their ratio gives the desired SNR gain.
\subsubsection{All-Inclusive Scheduling} In order to derive ${\sf E}(\rho_0)$ with $\rho_0$ defined in \eqref{aligned_power}, the distribution of $r_{\max} = \max_k \{r_1, r_2, \cdots, r_K\}$ is required which is provided as follows.
\begin{lemma}[Distribution of maximum distance]\label{lemma:2} \emph{The PDF of the maximum distance  $r_{\max}$ under the uniform user distribution in \eqref{pdf_r_k} is given by 
\begin{align}
f_{r_{\max}}(r) = \frac{2K}{R^{2K}} r^{2K -1}.
\end{align}
The result follows straightforwardly from  \eqref{pdf_r_k} and the proof is omitted. 
}
\end{lemma}

Using Lemma \ref{lemma:2}, the expected receive SNR for  all-inclusive scheduling is derived below.
\begin{lemma}[Expected receive SNR for all-inclusive scheduling]\label{prop:3} \emph{By employing all-inclusive scheduling, the resultant expected receive SNR is given by
\begin{align}
{\sf E}(\rho_0) = \frac{2K}{2K-\alpha} \frac{P_0}{M R^\alpha {\sf Ei}(g_{\sf th})}, \qquad  K>\frac{\alpha}{2}.
\end{align}}
\vspace{-2mm}
\proof
See Appendix \ref{app:prop:3}.
\endproof
\emph{
Since the path-loss exponent $\alpha \in [3,4]$ in practice, the requirement of $K>\frac{\alpha}{2}$ for the above result can be easily satisfied by having the number of edge devices $K > 2$.  
}
\end{lemma}

\subsubsection{Cell-Interior Scheduling} For cell-interior scheduling,  the following  result can be derived.
\begin{lemma}[Expected  receive SNR for cell-interior scheduling]\label{prop:4} \emph{By employing cell-interior scheduling, the resultant expected  receive SNR is given by
\begin{align}\label{eq:prop:4}
{\sf E}[\rho_0(R_{\sf in})] = \frac{c(R_{\sf in}) P_0}{M R_{\sf in}^\alpha {\sf Ei}(g_{\sf th})}, 
\end{align}
where $c(R_{\sf in})$ is a bounded scaling factor depending on $R_{\sf in}$ and $K$ with
\[c(R_{\sf in}) = \sum_{k = 2}^K \frac{2k}{2k - \alpha} \binom{K}{k} \l(\frac{R_{\sf in}^2}{R^2} \r)^k \l( 1-\frac{R_{\sf in}^2}{R^2} \r)^{K-k}.\] 
Particularly, for the typical case that $\alpha = 3$, we can show that $1 \leq c(R_{\sf in}) \leq 4$.}
\proof
See Appendix \ref{app:prop:4}.
\endproof
\end{lemma}

A direct comparison between Lemma \ref{prop:3} and \ref{prop:4} yields the SNR gain of cell-interior scheduling over the all-inclusive counterpart as shown below.

\begin{proposition}[SNR gain of cell-interior scheduling]\label{prop:G_SNR} \emph{Given the distance threshold $R_{\sf in}$, the cell-interior scheduling can attain the following receive SNR gain over the all-inclusive scheduling:
\begin{align}\label{Rx_power_gain}
G_{\sf SNR} = \frac{{\sf E}[\rho_0(R_{\sf in})]}{{\sf E}(\rho_0)} = a \l(\frac{R}{R_{\sf in}} \r)^\alpha,
\end{align}
where $a = \frac{2K-\alpha}{2K} c(R_{\sf in})$ is a bounded scaling factor, with $c(R_{\sf in})$ given in Lemma \ref{prop:4}.}
\end{proposition}
Note from Propositions \ref{prop:F_DAT} and \ref{prop:G_SNR} that both the fraction of exploited data and the SNR gain of the cell-interior scheduling are non-linear functions of the range ratio $\frac{R_{\sf in}}{R}$, but with different exponent scalings: the former is the square power law while the latter being a power law with the exponent equal to the path-loss exponent $\alpha$. 

\subsection{Reliability-Quantity Tradeoff}\label{sec:reliability_quantity_tradeoff}
Based on Proposition \ref{prop:F_DAT} and \ref{prop:G_SNR}, the mentioned tradeoff between update reliability and data quantity can be derived as follows.
\begin{proposition}[Reliability-quantity tradeoff]\label{prop:Q_R_tradeoff}
\emph{When cell-interior scheduling is employed, the tradeoff between the SNR gain and the fraction of exploited data for model training is given by 
\begin{align}
G_{\sf SNR} = a \l( \frac{1}{F_{\sf DAT}} \r)^{\frac{\alpha}{2}}
\end{align}
}
\end{proposition}

%\begin{Remark}[Reliability-quantity tradeoff]\label{remark:data_snr_tradeoff}
%\emph{
Proposition \ref{prop:Q_R_tradeoff} suggests that the path-loss component $\alpha$  plays a crucial role in determining how much SNR gain can be attained at the cost of losing a fraction $(1-F_{\sf DAT})$ of training data.
The larger the value of $\alpha$ is, the higher the cost is. Next, the result also provides a guideline for the selection of the distance threshold $R_{\sf in}$. Generally speaking, for large $\alpha$, the learning performance is more \emph{SNR-limited}.  It is thus desired to have a smaller $R_{\sf in}$ to alleviate the SNR penalty due to the scheduling of  cell-edge devices. In contrast, for small $\alpha$, the learning performance is limited by the size of training dataset, and thus \emph{data-limited}. Thereby, it is preferable to increase $R_{\sf in}$ to include more remote data at the edge to the training set with a degraded but acceptable receive SNR. 
%}
%\end{Remark}

We note that the optimal $R_{\sf in}$ that achieves the best learning accuracy is  challenging to derive  since how the said tradeoff affects the learning performance has  complex dependence on the data distribution, learning task and the learning model. Nevertheless, a general insight can be obtained is that, in the case of non-IID data distribution (see Section \ref{sec:simulation} for the definition), the learning performance tends to be more data-limited and thus a larger $R_{\sf in}$ is desired compared with the IID case.

\subsection{Coping with Data Deficiency}
The data deficiency of cell-interior scheduling, namely, the failure of exploiting cell-edge data, may lead to learning performance degradation. Two methods for addressing the issue are discussed in the sequel.

\subsubsection{High-mobility networks}
Consider the scenario where edge devices have high mobility and their locations change rapidly over time. In this scenario, the cell-interior scheduling is also known as \emph{opportunistic scheduling}. Given high mobility, the scheme can automatically cope with the data deficiency since an cell-edge device can enter the cell-interior in a subsequent communication round and be scheduled. 
By assuming the locations of all devices are i.i.d. over communication rounds,  we can quantify this fact as follows. 

\begin{proposition}
\emph{
In a high-mobility network with $K$ devices and given a training period consisting of $N_{\sf CR}$ communication rounds, the probability that all distributed data is exploited for learning is given by
\begin{align}
p_{\sf all} &= \l(1 - (1 - p_{\sf in})^{N_{\sf CR}} \r)^K \\
& \approx 1 - K ( 1 - p_{\sf in})^{N_{\sf CR}}, \qquad N_{\sf CR} \to \infty,
\end{align}
}
\end{proposition}
where $p_{\sf in} = \l( \frac{R_{\sf in}}{R} \r)^2$ denotes the probability that a device lies in the cell-interior. 
\proof
The proof is straightforward by noting that the event that all data is exploited during $N_{\sf CR}$ communication rounds is equivalent to that all devices are \emph{ever} in the cell-interior in $N_{\sf CR}$ communication rounds. The detailed derivation is omitted  for brevity.
\endproof
Note that, as $N_{\sf CR}$ increases, the probability $p_{\sf all}$ approaches to 1 at an exponential rate. This justifies our claim that the opportunistic (cell-interior) scheduling can efficiently cope with the data deficiency issue by simply increasing $N_{\sf CR}$ to exploit the device mobility. 

\subsubsection{Low-mobility networks}
%Purely scheduling the cell-interior user may lead to a high effective update-SNR but may suffer from exploiting only a small portion of distributed data. This is the case where the edge devices remain static during the training. 
However, in the low-mobility networks, the fraction of exploited data by cell-interior scheduling remains unchanged over  communication rounds.
For this reason, we propose the following alternating-scheduling scheme to exploit the cell-edge data.

\begin{framed}
\vspace{-10pt} 
\begin{scheme}[Edge-Interior Alternating scheduling]\label{scheme:2}
\emph{The edge server alternates between  cell-interior scheduling (Scheme 1) and all-inclusive scheduling.}
\end{scheme}
\vspace{-10pt} 
\end{framed}

By alternating cell-interior and all-inclusive scheduling, the current scheme strike a balance between the advantages/disadvantages of the two sub-schemes. In particular, alternating scheme can exploit all data for learning while achieving an effective receive SNR averaging those of the two sub-schemes. The alternating frequency between the two modes could be  optimized to balance the reliability-quantity tradeoff for  improving learning performance. As a result, the alternating scheduling can outperform both the cell-interior and all-inclusive schedulings.

\section{Latency Analysis: Broadband Analog  Aggregation v.s. Broadband Digital  Aggregation}

\begin{figure}[tt]
  \centering
    \subfigure[Broadband analog aggregation by AirComp]{\label{subfig:analog}\includegraphics[width=0.44\textwidth]{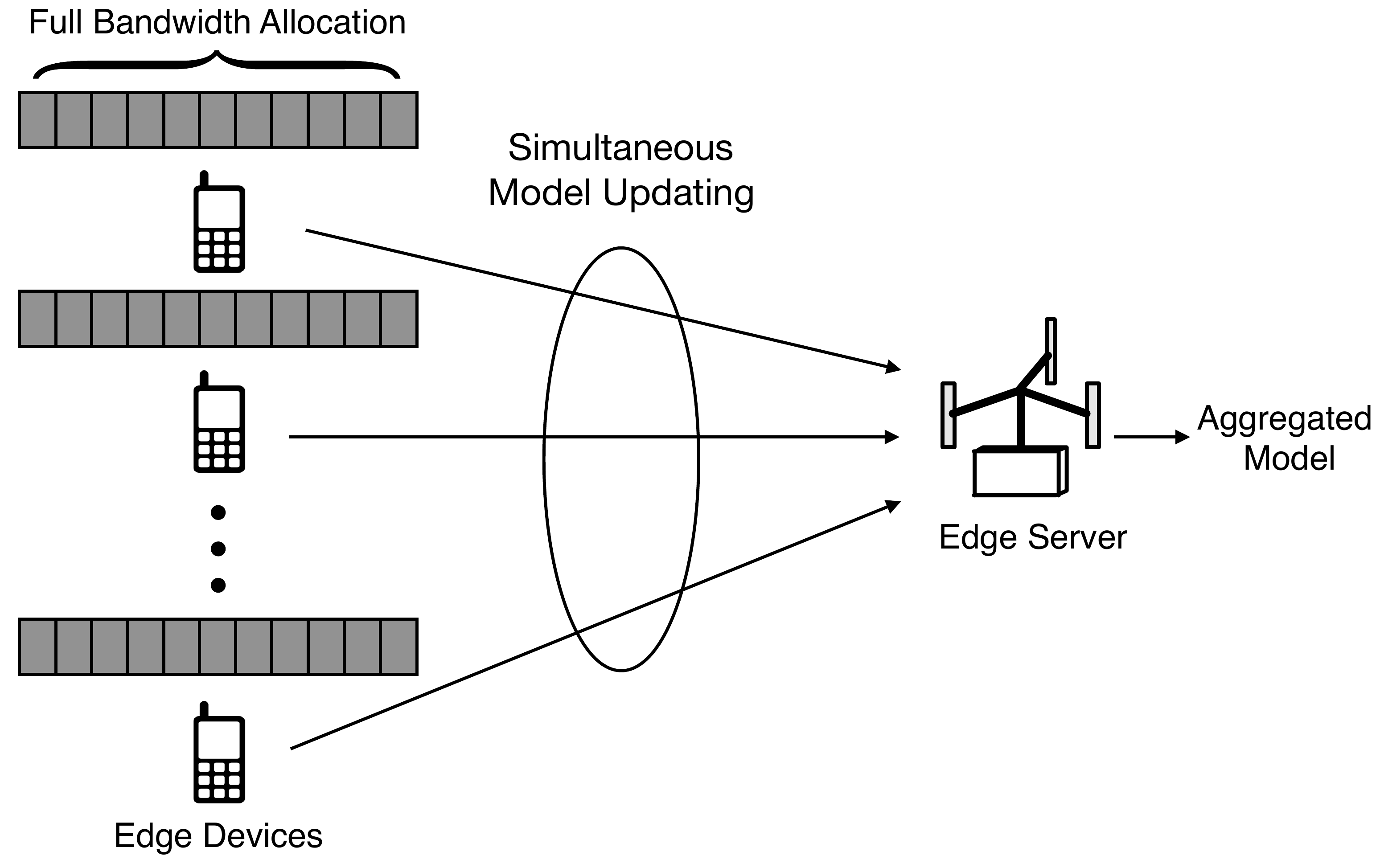}}
      \hspace{0.2in}
  \subfigure[Broadband digital  aggregation by OFDMA]{\label{subfig:digital}\includegraphics[width=0.51\textwidth]{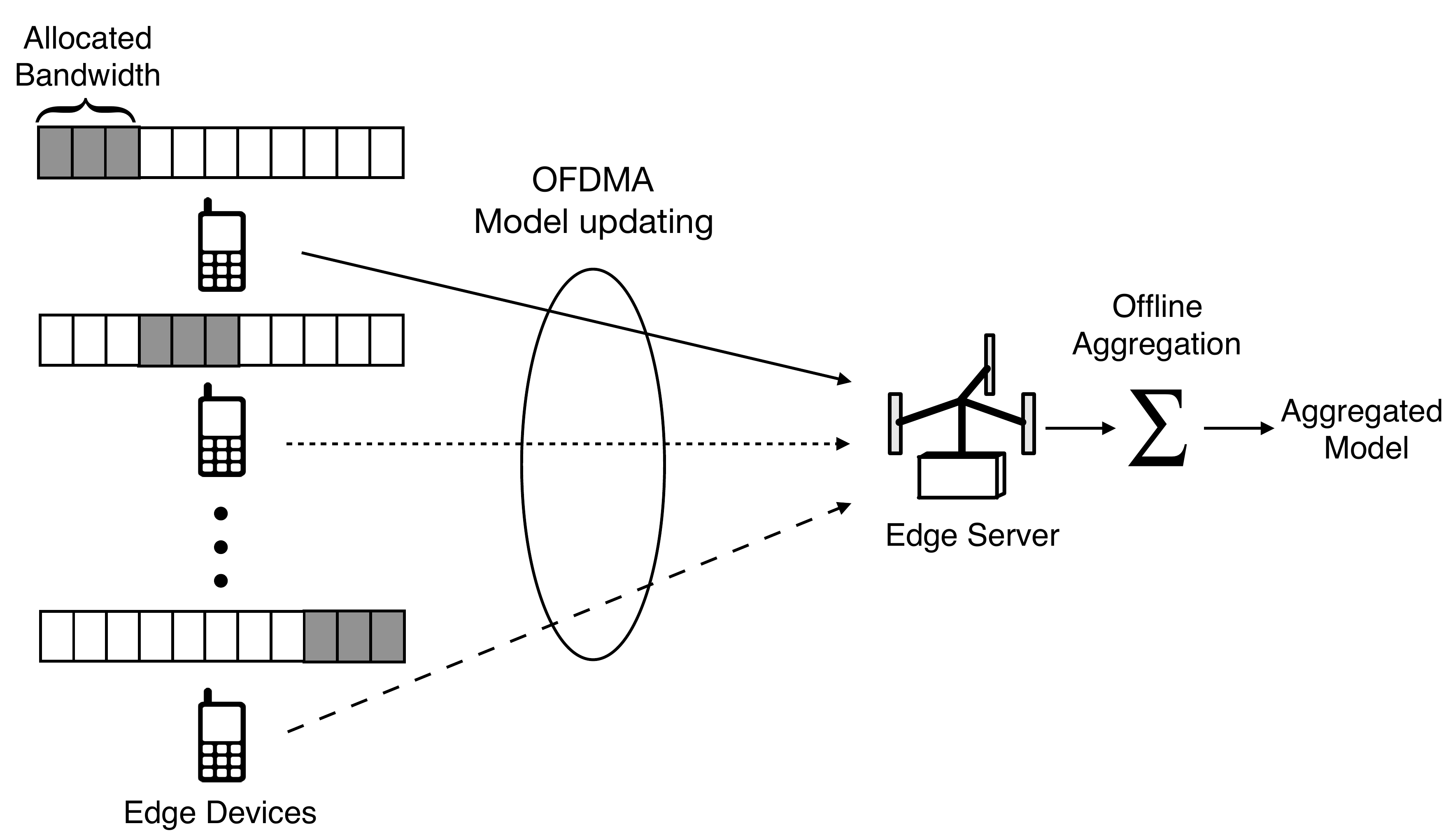}}
   \vspace{3mm}
  \caption{Broadband analog aggregation versus broadband digital aggregation.}
  \label{Fig:analog_agg_vs_digital_agg}
\end{figure}

The key advantage of the proposed BBA w.r.t. the conventional digital  OFDMA is the significant reduction in communication latency. As illustrated in Fig. \ref{Fig:analog_agg_vs_digital_agg}, the fundamental reason for the latency reduction is the difference in  how the two schemes allocate the spectrum to devices. BAA allows the complete reuse of the whole bandwidth to exploit ``interference'' for direct aggregation. OFDMA orthogonalizes the bandwidth allocation to avoid interference for offering reliable communication. 
As a result, the bandwidth per device reduces with the number of devices. 
In this section, we analyze the latency reduction of BAA w.r.t. OFDMA.

\subsection{Latency Analysis of Broadband Analog Aggregation}
For BAA, each model parameter is amplitude-modulated to a single analog symbol and each sub-channel is dedicated for a single parameter transmission. Thus, to upload a model update of dimension $q$, the total number of analog symbols to be transmitted is calculated as
\begin{align}\label{analog_symbol} (\text{Analog symbols for one update}) \qquad
D_{\sf ana} = q \;(\text{symbols}).
\end{align} 
Since all devices transmit their model-updates simultaneously using all  available sub-channels. One can easily derive that the number of OFDM symbols required for transmitting the whole update is equal to $\frac{q}{M}$. Thus, we obtain the following  result.

\begin{proposition}[Latency for BAA]\label{prop:ana_latency}
\emph{The latency per communication round for  BAA  is given by
\begin{align}\label{Analog_latency}
T_{\sf ana} = \frac{q}{M} T_s,
\end{align}
where $T_s$ is the symbol duration of an OFDM symbol.
}
\end{proposition} 

Two key observations can be made from the result in \eqref{Analog_latency} as follows:
\begin{itemize}
\item Due to the complete reuse of radio resource (e.g., time and frequency) among devices, the resultant latency is thus independent of the number of accessing devices, making it particularly promising in dense edge-learning network.

\item The latency of the analog  aggregation is a deterministic value independent of the channel realizations, which is in contrast to the digital counterpart whose latency is a random variable due to the channel-dependent transmission rate as will be shown in \eqref{Tx_rate_instant}.
\end{itemize}

\subsection{Latency Analysis of Broadband Digital  Aggregation} 
For  broadband digital  aggregation (OFDMA), each parameter is  first quantized into a fixed number of bits, denoted as $Q$. Then, for a device to upload a model update of dimension $q$, the total data amount to be transmitted is  
\begin{align}\label{data_amount} (\text{Data amount for one update}) \qquad
D_{\sf dig} = qQ \;(\text{bits}).
\end{align} 
During update aggregation, all $K$ edge devises communicate with the edge server based on OFDMA to avoid inter-device interference. For simplicity, we assume that the total available bandwidth is evenly divided and assigned to $K$ devices, so each device uploads its local model via an equal portion of allocated sub-channels [see Fig. \ref{subfig:digital}]. Thus the number of sub-channels allocated to device $k$ is given by
\begin{align} (\text{Allocated sub-channels per device}) \qquad
M_k = \frac{M}{K}.
\end{align} 
Thus the received signals from device $k$ on the $m$-th sub-channel can be rewritten from \eqref{channel_model} as
\begin{align}\label{digital_system_model}
y_k^{(m)} = r_k^{-\frac{\alpha}{2}}h_k^{(m)} p_k^{(m)} x_k^{(m)} + z^{(m)}.
\end{align}
where the notations follow those in \eqref{channel_model} with the index of OFDM symbol  omitted to avoid heavy notation. $x_k$ denotes the quantized version of the model-update parameter. Since only a fraction of spectrum is used by each device, the power constraint in \eqref{power_constraint2} is modified as follows:
\begin{align}\label{power_constraint3} (\text{Power constraint}) \qquad
{\sf E} \l[  | p_k^{(m)}(h_k^{(m)})|^2 \r]  &\leq \frac{K P_0}{M}, \qquad \forall k.
\end{align} 

%Starting from \eqref{digital_system_model}, it is noted that the practical transmission rate of the system is determined by the power control policy and the modulation scheme adopted. 

In order to derive the model updating latency, the transmission rate of the system is needed.
To this end, we consider the practical adaptive QAM modulation scheme \cite{goldsmith1997variable}.
It is well known that the optimal power control for such a scheme follows ``water-filling'' over channel realizations. The resultant transmission rate has no closed-form, making latency analysis intractable. The difficulty can be overcome by considering the sub-optimal power control, truncated channel inversion in \eqref{truncated_channel_inv}. Then based on the result from \cite{goldsmith1997variable}, given a target \emph{bit error rate} (BER), the resultant transmission rate for device $k$ on sub-channel $m$ is:
\begin{align}\label{Tx_rate_instant} (\text{Instantaneous transmission rate}) \qquad
 R_k^{(m)} = \l\{\begin{aligned} 
& B_{\sf sub} \log_2 \l( 1 + \frac{-1.5 \rho_k}{\ln(5\text{BER})} \r), && |h_k^{(m)}|^2 \geq g_{\sf th} \\
&0, &&  |h_k^{(m)}|^2 < g_{\sf th},
\end{aligned}
\r.
\end{align}
where $B_{\sf sub} = \frac{B}{M}$ denotes the sub-carrier spacing in the OFDM system and the receive SNR $\rho_k$ can be easily derived by substituting \eqref{truncated_channel_inv} into \eqref{power_constraint3}:
\begin{align}\label{eq_dig_rx_snr}
\rho_k = \frac{KP_0}{M r_k^\alpha {\sf Ei}(g_{\sf th})}. 
\end{align}
By taking expectation of \eqref{Tx_rate_instant} w.r.t. sub-channel coefficient $h_k^{(m)}$ and summing over all the allocated sub-channel indeces $\{m\}_{m=1}^{M_k}$, the expected sum transmission rate for device $k$ can be computed as follows:
 \begin{align}\label{Tx_rate_expected}
 R_k = {\sf E}\l( \sum_{m = 1}^{M_k} R_k^{(m)} \r) = M_k B_{\sf sub} \log_2 \l( 1 + \frac{-1.5 \rho_k}{\ln(5\text{BER})} \r) {\sf Pr} (|h_k^{(m)}|^2 \geq g_{\sf th}).
 \end{align}
Based on the Rayleigh-fading channel model, we have ${\sf Pr} (|h_k^{(m)}|^2 > g_{\sf th}) =  \exp(-g_{\sf th})$. Thus \eqref{Tx_rate_expected} can be explicitly given by
\begin{align}\label{Tx_rate2}  (\text{Expected transmission rate}) \qquad
R_k =  M_k B_{\sf sub} \log_2 \l( 1 + \frac{-1.5 K P_0}{\ln(5\text{BER})M r_k^\alpha {\sf Ei}(g_{\sf th})} \r) \exp(-g_{\sf th}).
\end{align}
Given \eqref{data_amount} and \eqref{Tx_rate2}, we derive the expected update communication latency for device $k$, denoted by $T_k$, as follows.
\begin{align} \qquad
T_k = \frac{D_{\sf dig}}{R_k} = \frac{KqQ}{M B_{\sf sub} \log_2\l(1 + \frac{-1.5 K P_0}{\ln(5\text{BER}) M r_k^\alpha {\sf Ei}(g_{\sf th})}\r)\exp(-g_{\sf th})}.
\end{align}

Since the model aggregation is performed offline by the edge server after all local models are reliably received, the communication latency is determined by that of the slowest device, which is known as the \emph{straggler effect}. 
Accordingly, we can establish the main result in the current sub-section as follows.
\begin{proposition}[Expected latency for broadband digital aggregation]\label{prop:digital_latency}
\emph{The expected latency per communication round for broadband digital model aggregation  is given by
\begin{align}\label{Digtial_latency}
T_{\sf dig} = \max_k \{T_k\} = \frac{KqQ}{M \log_2\l(1 + \frac{-1.5 K P_0}{\ln(5\text{BER}) M r_{\max}^\alpha {\sf Ei}(g_{\sf th})}\r)\exp(-g_{\sf th})} T_s,
\end{align}
where $r_{\max} = \max_k \{r_1, r_2, \cdots, r_K\}$ denotes the distance to the furthest user, and $T_s = \frac{1}{B_{\sf sub}}$ is the symbol duration of an OFDM symbol.}
\end{proposition} 

Several key observations can be made from \eqref{Digtial_latency} as summarized below:
\begin{itemize}
\item The latency of the scheme approximately linearly scales with the number of accessing devices $K$.  
\item Due to the straggler effect, the latency of the scheme is bottlenecked by the distance to the furthest user in the network $r_{\sf max}$. The level of  latency penalty for scheduling a far-away user is determined by the path-loss exponent $\alpha$.
\item The latency can be controlled by the target  BER. Lower BER can accelerate the update aggregation but at a cost of degraded update-reliability and vice versa. 
\end{itemize}

%During the federated model training, in each communication round, local models trained at edge devices (using e.g., stochastic gradient descend) are transmitted and aggregated at the edge server over a shared broadband channel that consists of $M=1000$ orthogonal sub-channels. Two multiple access schemes, namely the conventional OFDMA and the proposed AirComp, are compared. They mainly differ in how the available sub-channels are shared. For OFDMA, the 1000 sub-channels are evenly allocated to the K edge devices, so each device uploads its local model using only fractional bandwidth that reduces as K grows. Model averaging is performed by the edge server after all local models are reliably received, and thus the communication latency is determined by the slowest device. In contrast, the AirComp scheme allows every device to use the full bandwidth so as to exploit the ``interference'' for direct model averaging over the air. The latency of AirComp is thus independent of the number of accessing devices. 

%\subsection{Implementation details}
%For AirComp, model parameters are analog-modulated and each sub-band is dedicated for single-parameter transmission; truncated-channel inversion (power control) under the transmit-power constraint is used to tackle the channel fading. For OFDMA, model parameters are first quantized into a bit sequence (16-bit per parameter). Then adaptive MQAM modulation is adopted to adapt the data rate to the channel condition such that the spectrum efficiency is maximized while the target bit-error-rate of $10^{-3}$ is maintained.

\subsection{Latency Comparison between Analog and Digital Aggregation}
Combining Propositions   \ref{prop:ana_latency} and \ref{prop:digital_latency}, we are ready to derive the latency-reduction ratio of BAA w.r.t. the digital counterpart, defined as $\gamma = \frac{T_{\sf dig}}{T_{\sf ana}}$, as follows.

\begin{proposition}[Latency reduction]\label{prop:latency_gain}\emph{
The latency-reduction ratio of the BAA over its digital counterpart, is given by
\begin{align}
\gamma = \frac{KQ}{ \log_2\l(1 + \frac{-1.5 K P_0}{\ln(5\text{BER}) M r_{\max}^\alpha {\sf Ei}(g_{\sf th})}\r)\exp(-g_{\sf th})}.
\end{align} 
}
\end{proposition}

Based on Proposition \ref{prop:latency_gain}, the following insights can be derived.
\begin{itemize} 
\item The latency-reduction ratio  scales linearly with the quantization resolution $Q$ and approximately linearly  with the number of devices $K$. More precisely, we have the following scaling law w.r.t. $K$:
\begin{align}
\gamma = O\l(\frac{K}{\log_2K}\r), \qquad K \to \infty.
\end{align}
\item The latency-reduction ratio can keep increasing unboundedly as $r_{\sf \max} \to \infty$ and the increasing rate  depends on the path-loss exponent $\alpha$.

\item The latency-reduction ratio is a monotone decreasing function of the target BER. Particularly, as $\text{BER} \to 0$, the ratio grows unboundedly, i.e., $\gamma \to \infty$, since no practical modulation scheme can achieve zero BER.

\item For the digital scheme, the power-cutoff threshold has double effects on the latency-reduction ratio via affecting the receive SNR and channel-cutoff probability. On one hand, increasing $g_{\sf th}$ leads to a higher receive SNR of the digital scheme as reflected in \eqref{eq_dig_rx_snr}, which reduces the latency-reduction ratio. On the other hand, a large $g_{\sf th}$ incurs a high channel-cutoff probability, that reduces the expected transmission rate of the digital scheme [see \eqref{Tx_rate2}] and thus increases the latency-reduction ratio.   
\end{itemize}

A comprehensive comparison between analog and digital aggregation is summarized in Table \ref{summary:table1}.

\begin{table}[tt]
\centering
\caption{Comparison between analog and digital model aggregation.}
\vspace{2mm}
\begin{tabular}{|p{4.4cm}|p{6.6cm}|p{5.7cm}|}
\hline
{} & \bf{Broadband analog aggregation} & \bf{Broadband digital aggregation} \\
\hline
\bf{Effect of channel condition} &  Receive SNR and Truncation ratio [see \eqref{aligned_power} \& \eqref{eq:truncation_ratio}] &  Transmission rate [see \eqref{Tx_rate_instant}] \\ 
\hline
\bf{Distance dependency}   & Receive SNR depends on furthest user [see \eqref{aligned_power}] &  Latency depends on furthest user [see \eqref{Digtial_latency}] \\ 
\hline
\bf{Latency scaling with \# of devices}   &  Independent [see \eqref{Analog_latency}] & Approximately linear scaling [see \eqref{Digtial_latency}] \\ 
\hline
\bf{Update reliability guarantee}   &  Loose guarantee by scheduling [see \eqref{Rx_power_gain}] &  Strict guarantee specified by target BER \\ 
\hline
\end{tabular}
\label{summary:table1}
\vspace{-4mm}
\end{table}

\section{Experimental Results} \label{sec:simulation}
\subsection{Experiment Settings} \label{simulation1}
Consider a FEEL system with one edge server and $K=200$ edge devices. The simulation parameters are set as follows unless specified otherwise. The cell radius is $R = 100$, the path loss exponent $\alpha = 3$, the number of sub-channels $M=1000$, the average transmit power constraint per device $P_0 = 0.1 (W)$, and the channel noise variance $N_0 = -80$ {dBm}.

\begin{figure}[tt]
\centering
\includegraphics[width=18cm]{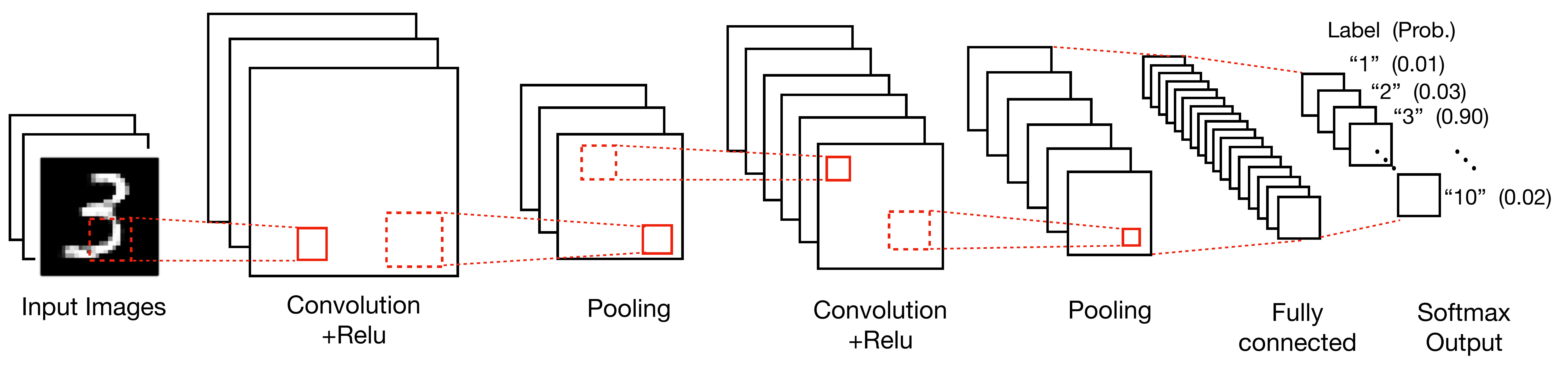}
\caption{Architecture of the convolutional neural network used in experiments.}
\label{Fig:convolutional_neural_network}
\end{figure}

For exposition, we consider the learning task of handwritten-digit recognition using the well-known MNIST dataset that consists of 10 categories ranging from digit ``0'' to ``9'' and a total of 60000 labeled training data samples. To simulate the distributed mobile data, we consider two types of data partitions, i.e., the {\bf IID} setting and {\bf non-IID} one. For the former setting, we randomly partition the training samples into 200 equal shares, each of which is assigned to one particular device. While for the latter setting, we first sort the data by digit label, divide it into $400$ shards of size $150$, and assign each of 200 clients 2 shards.
As illustrated in Fig. \ref{Fig:convolutional_neural_network}, the classifier model is implemented using a 6-layer convolutional neural network (CNN) that consists of two 5x5 convolution layers with ReLu activation  (the first with 32 channels, the second with 64), each followed with 2x2 max pooling, a fully connected layer with 512 units and ReLu activation, and a final softmax output layer (582,026 parameters in total).

%XXX: List of simulations:
%
%Can combine 1 and 2 in the same figures? 
%
%1. Show the necessity of user scheduling, the learning performance first increases with the number of users and then decreases after a certain point.
%
%2. Show the performance of the cell-interior scheduling for high mobility scenario.
%Benchmark schemes: cell-interior scheduling and all-inclusive scheme. 
%
%3. Show the performance of the alternating scheduling for static scenario. 
%Benchmark schemes: cell-interior scheduling and all-inclusive scheme. 
%
%4. Latency gain over digital scheme. Show IID and NIID data partitions's learning performance.  

\vspace{-4mm}
\subsection{Tradeoff in User Scheduling}

 \begin{figure}[tt]
  \centering
  \subfigure[$\alpha = 2.5$]{\label{subfig:alpha_2.5}\includegraphics[width=0.48\textwidth]{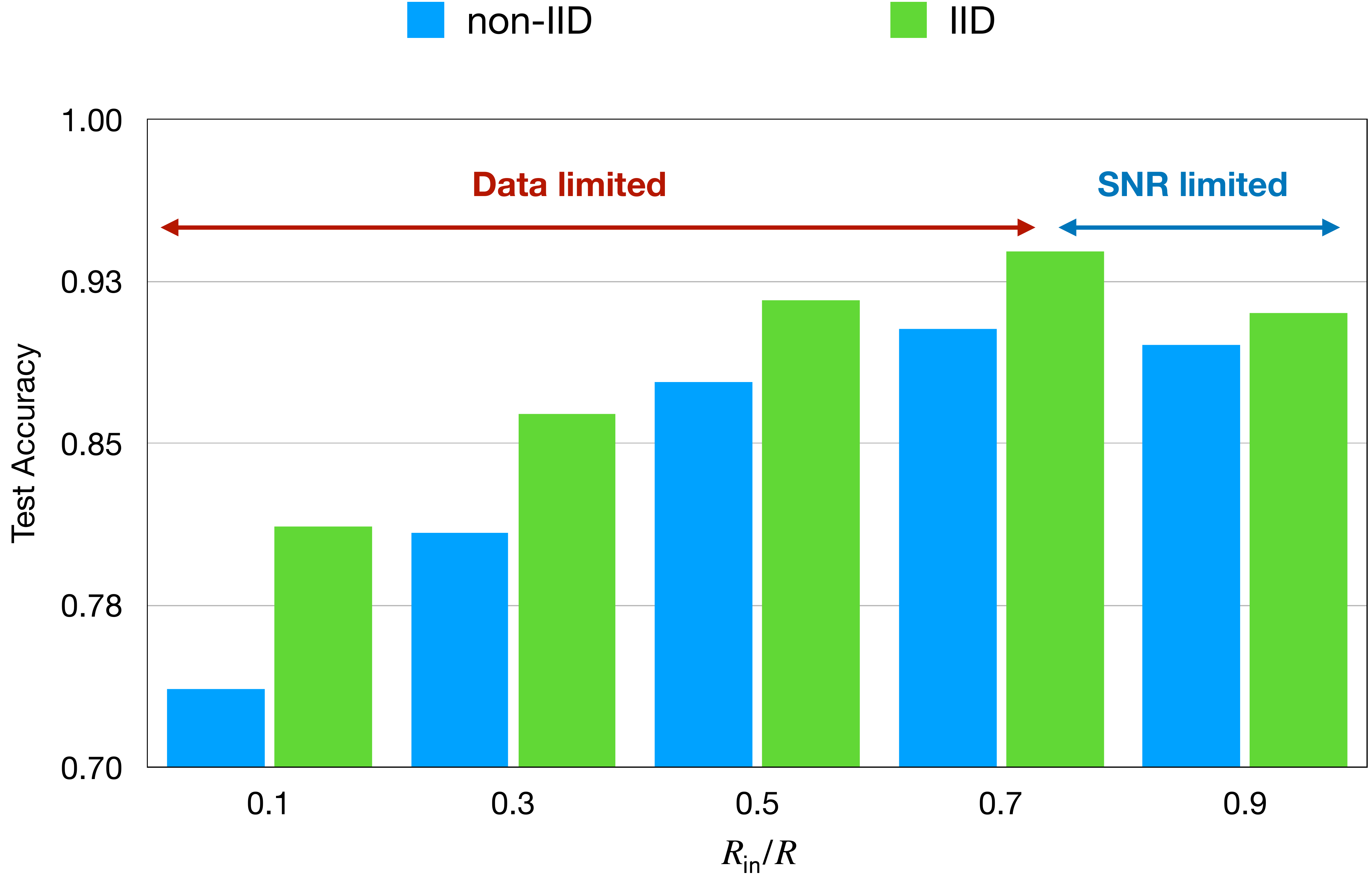}}
  \hspace{0.05in}
  \subfigure[$\alpha = 3$]{\label{subfig:alpha_3}\includegraphics[width=0.48\textwidth]{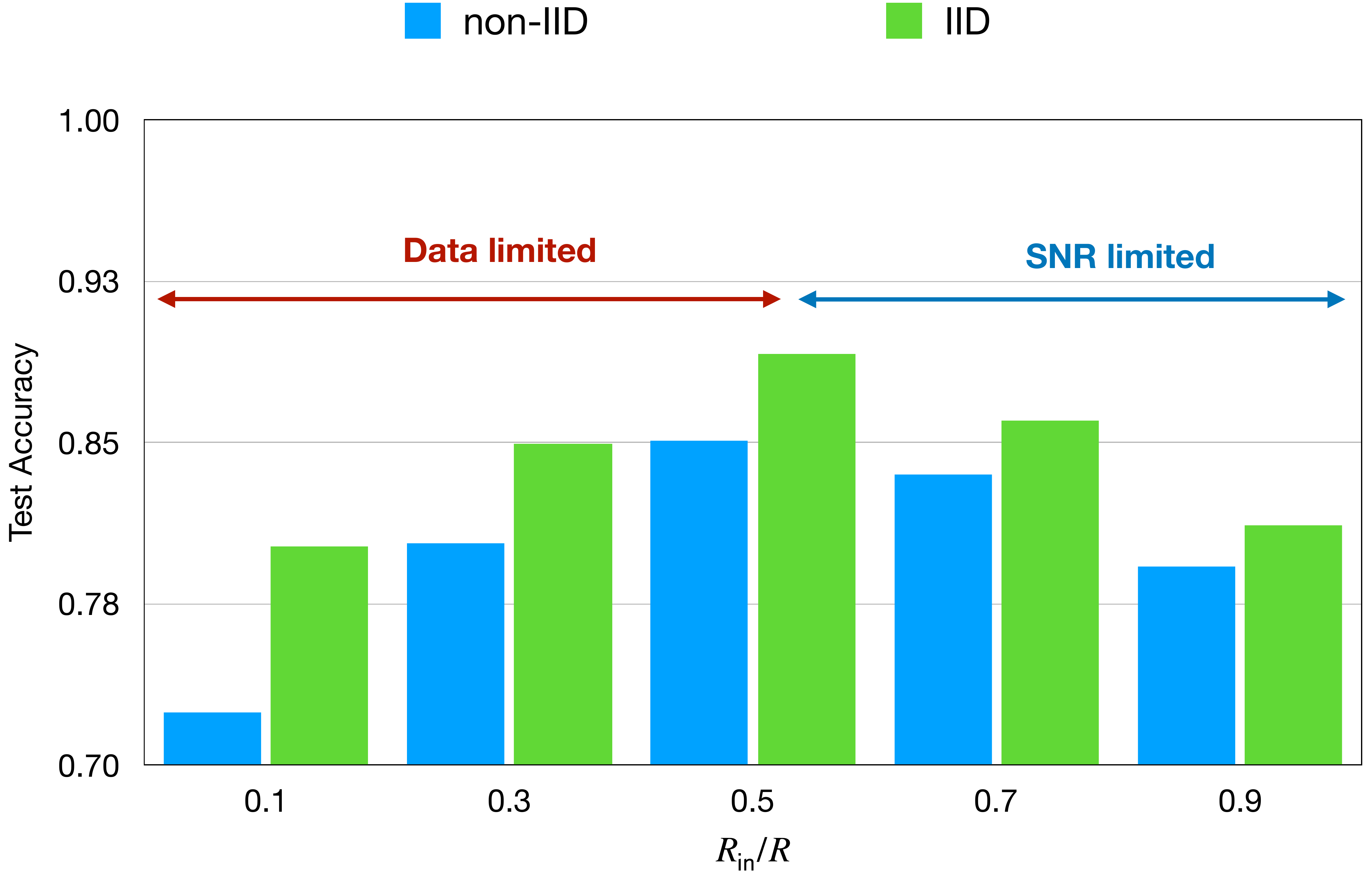}}
    \hspace{0.05in}
  \subfigure[$\alpha = 3.5$]{\label{subfig:alpha_3.5}\includegraphics[width=0.48\textwidth]{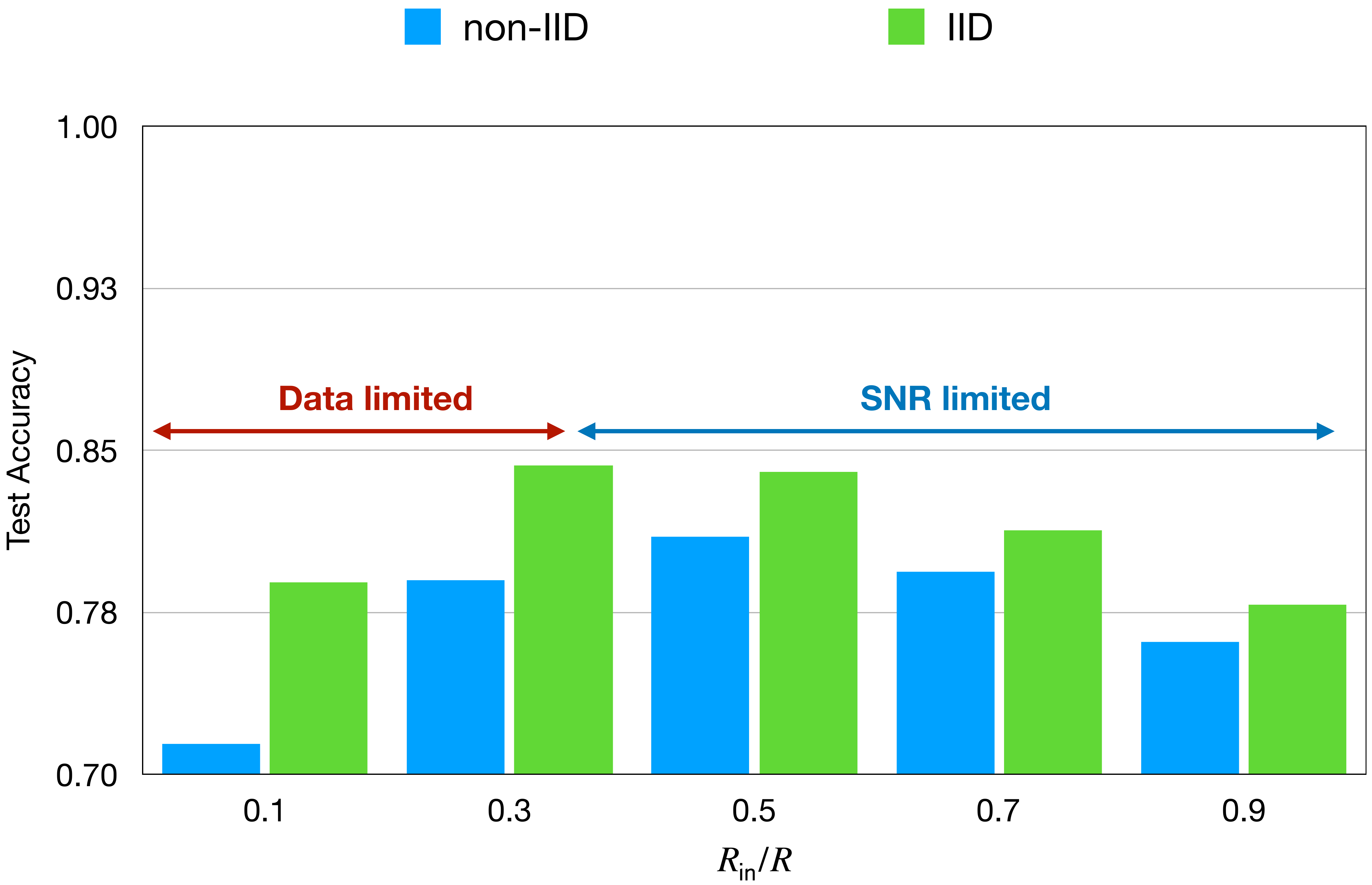}}
  \vspace{3mm}
  \caption{Test accuracy versus distance threshold in cell-interior scheduling.}
  \label{Fig:accuracy_vs_range}
  \vspace{-4mm}
\end{figure}

The tradeoff inherent in the user scheduling problem is first shown in Fig. \ref{Fig:accuracy_vs_range}.
 Consider the cell-interior scheduling in Scheme \ref{scheme:1}. The bar-figure showing the ultimate test accuracy of the learned model against the  selection of the normalized distance threshold $R_{\sf in}/R$ is plotted under varying values of the path-loss component $\alpha$. Both the cases of IID and non-IID data-partition are experimented. It can be observed from all plots that, as the more devices included in the aggregation by increasing $R_{\sf in}$, the test accuracy first increases then decreases after a certain point, passing through a data-limited regime towards a SNR-limited regime. The phenomenon verifies the existence of the said reliability-quantity tradeoff in user scheduling. In addition, as the path-loss exponent increases, the learning performance is found to be more suffered from SNR-limited than data-limited, suggesting a decreasing choice of $R_{\sf in}$ to reduce the SNR penalty due to the scheduling of cell-edge devices. Last, it is also noted that the non-IID setting is in general more data-hungry than the IID one, thereby preferring a higher value of $R_{\sf in}$ even when the path-loss exponent is high [see Fig. \ref{subfig:alpha_3.5}]. The observations align with our previous discussions in Section \ref{sec:reliability_quantity_tradeoff}.

% \begin{figure*}[tt]
%  \centering
%  \subfigure[i.i.d. data partition]{\label{subfig:dl_mc}\includegraphics[width=0.42\textwidth]{effect_of_num_user_iid.pdf}}
%  \hspace{0.35in}
%  \subfigure[non-i.i.d. data partition]{\label{subfig:ul_ac}\includegraphics[width=0.42\textwidth]{effect_of_num_user_niid.pdf}}
%  \vspace{3mm}
%  \caption{Effect of the number included devices in high mobility scenario with i.i.d./non-i.i.d. data partition.}
%  \label{Fig:effect_num_user}
%\end{figure*}

\vspace{-4mm}
\subsection{Performance Comparison between Different Scheduling Schemes}
 \begin{figure}[tt]
  \centering
  \subfigure[non-IID data distribution]{\label{subfig:niid_2}\includegraphics[width=0.48\textwidth]{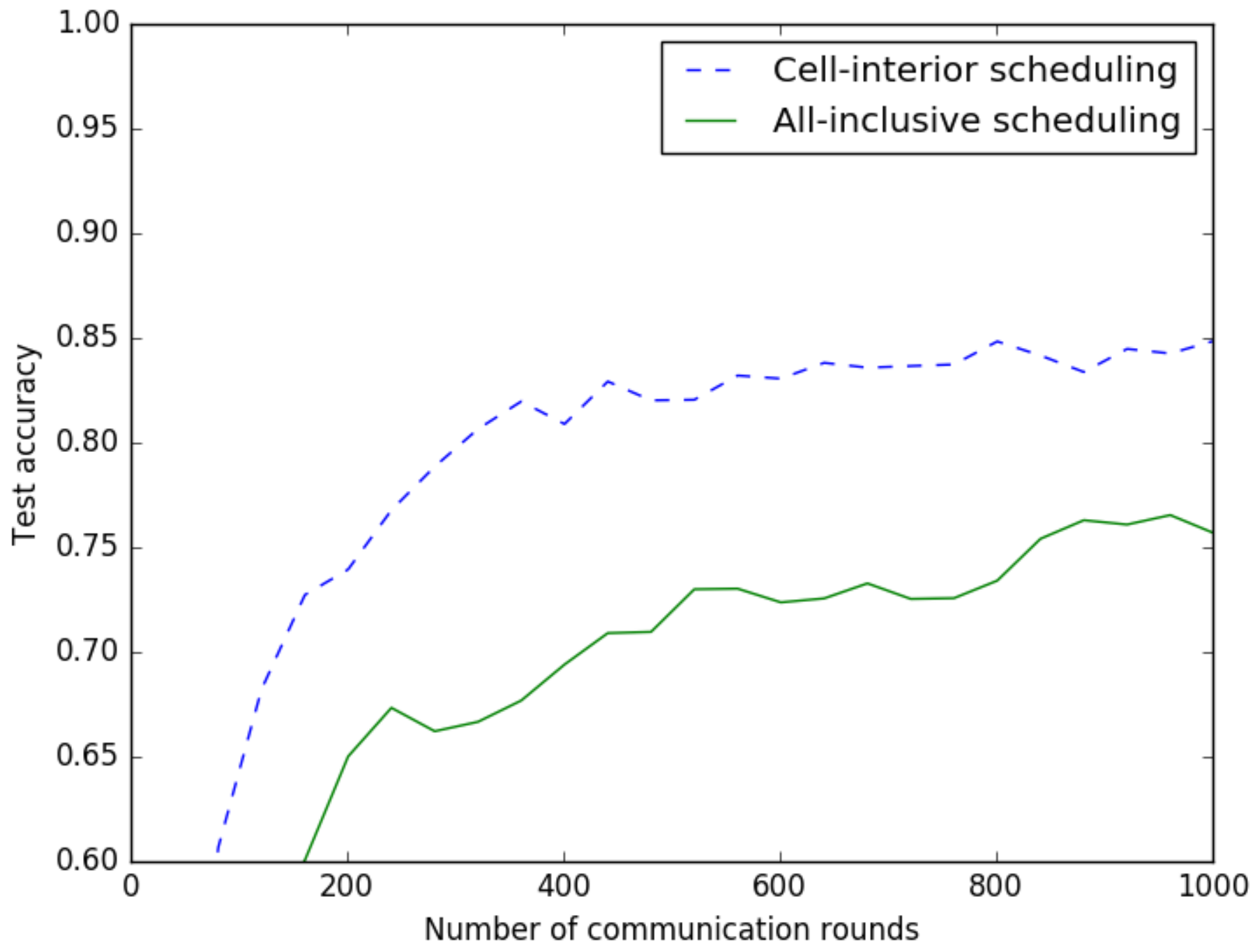}}
  \hspace{0.05in}
  \subfigure[IID data distribution]{\label{subfig:iid_2}\includegraphics[width=0.48\textwidth]{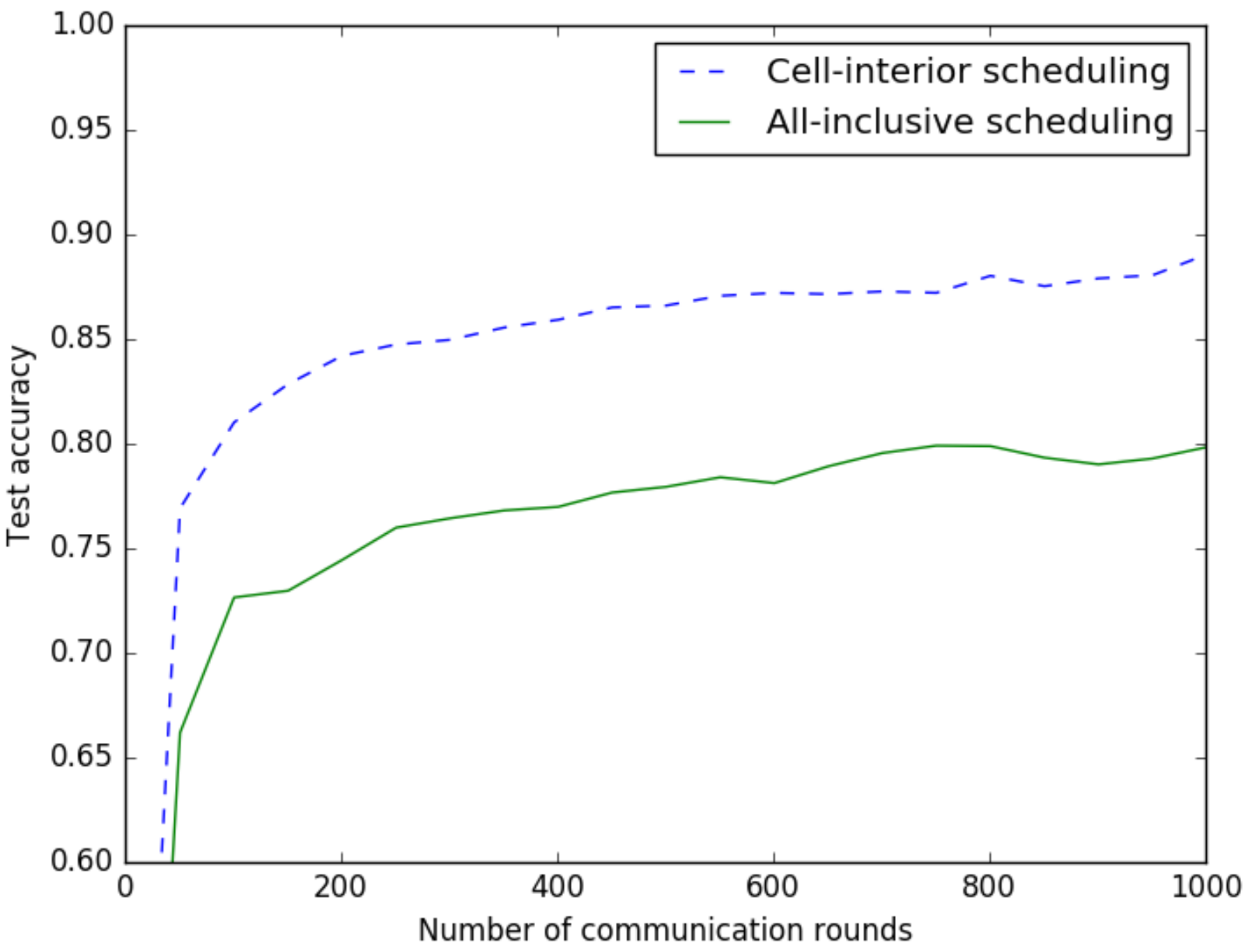}}
  \vspace{3mm}
  \caption{Performance comparison between different scheduling schemes in the high-mobility scenario.}
  \label{Fig:accuracy_vs_CR_mobililty}
    \vspace{-3mm}
\end{figure}

 \begin{figure}[tt]
  \centering
  \subfigure[non-IID data distribution]{\label{subfig:niid_1}\includegraphics[width=0.485\textwidth]{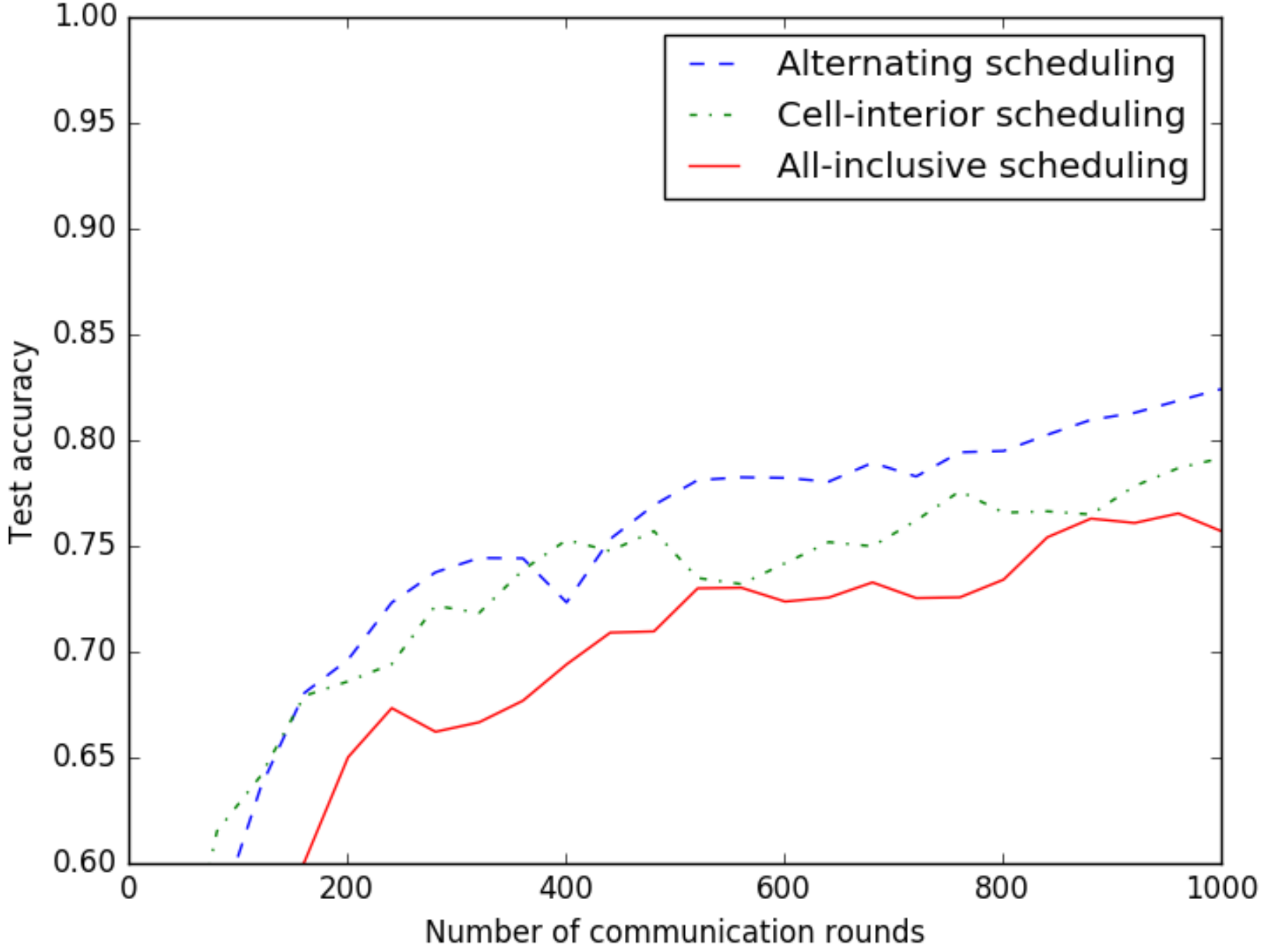}}
  \hspace{0.05in}
  \subfigure[IID data distribution]{\label{subfig:iid_1}\includegraphics[width=0.48\textwidth]{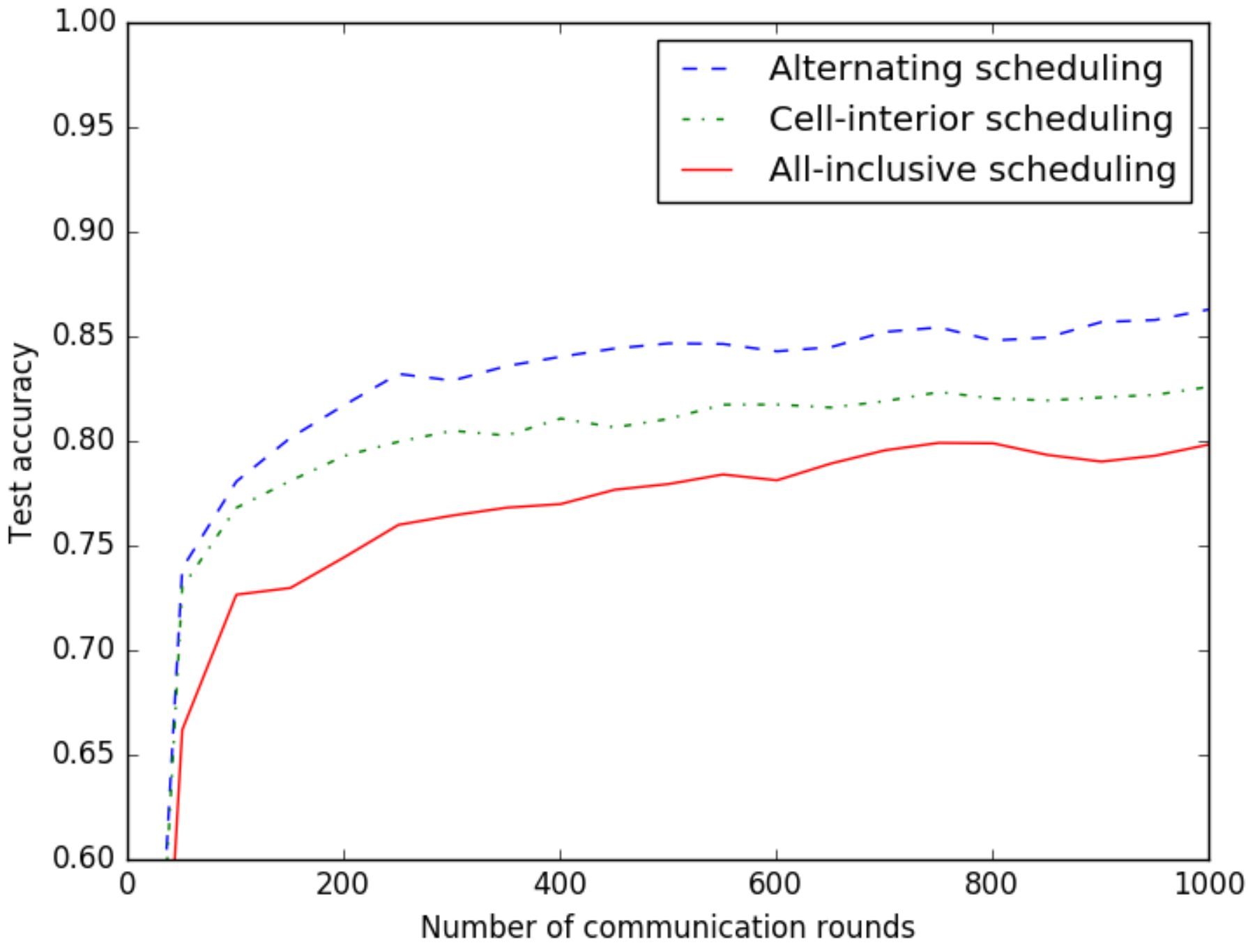}}
  \vspace{3mm}
  \caption{Performance comparison between different scheduling schemes in the low-mobility scenario.}
  \label{Fig:accuracy_vs_CR_static}
    \vspace{-4mm}
\end{figure}

The performance of the developed cell-interior scheduling scheme is evaluated in Fig. \ref{Fig:accuracy_vs_CR_mobililty} and \ref{Fig:accuracy_vs_CR_static}, targeting the high-mobility and low-mobility networks, respectively. The difference between the two scenarios is that, in the former setting, the devices lying within the cell-interior change rapidly over communication rounds, while in the latter case, the device locations remain unchanged throughout the entire learning process. For all curves, the distance threshold $R_{\sf in}$ is optimized numerically for the best test accuracy. It is observed that the cell-interior scheduling outperforms the naive all-inclusive scheme by a remarkable gap in the high-mobility scenario where the learning performance is more SNR-limited. On the other hand, in the low-mobility scenario where the cell-interior scheduling suffers from data-deficiency, the proposed alternating scheduling scheme can enhance the learning performance further by occasionally exploiting  the data in the cell-edge devices.

  \begin{figure}[tt]
  \centering
  \subfigure[Test accuracy]{\label{subfig:accuracy}\includegraphics[width=0.48\textwidth]{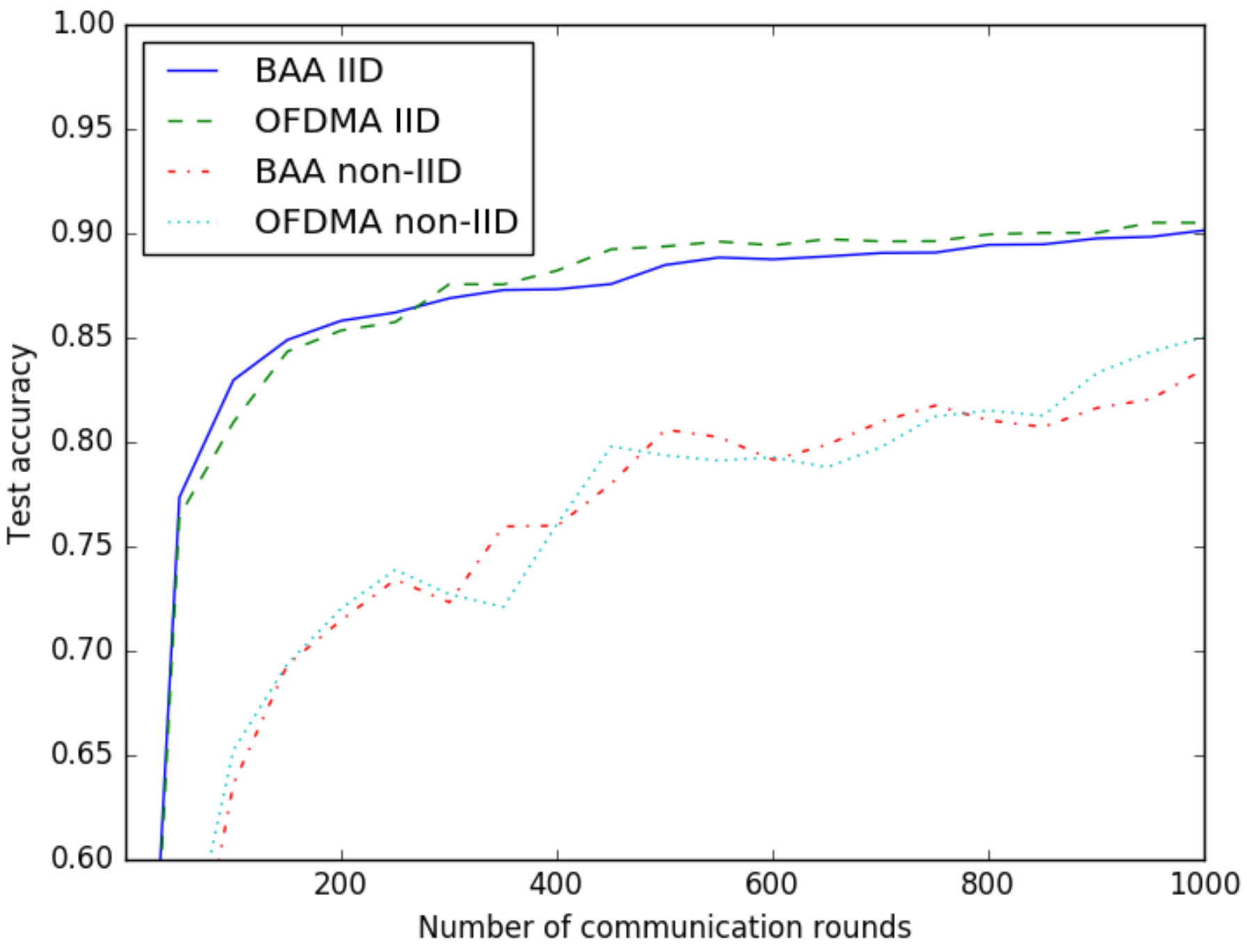}}
  \hspace{0.05in}
  \subfigure[Communication latency]{\label{subfig:latency}\includegraphics[width=0.47\textwidth]{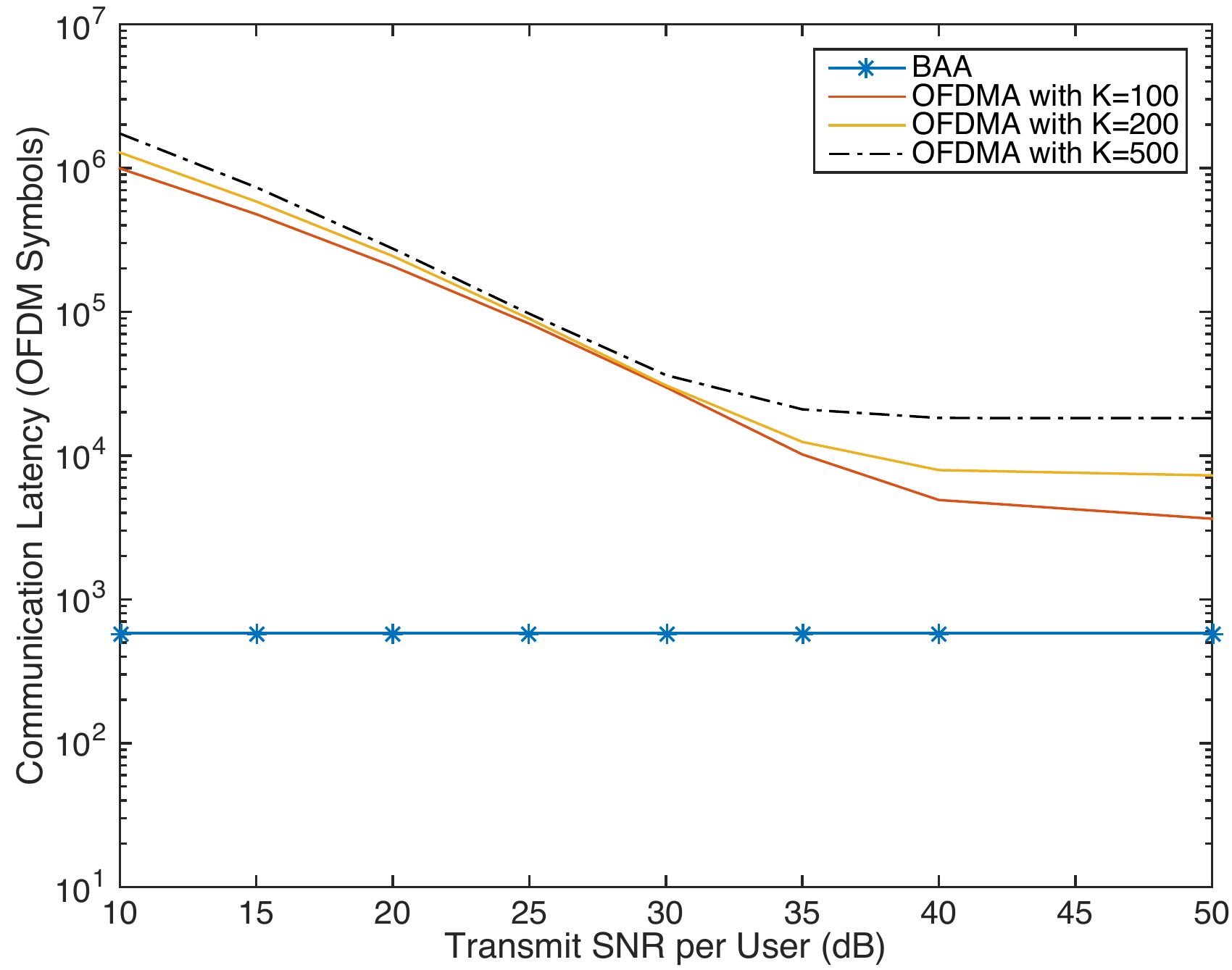}}
  \vspace{3mm}
  \caption{Performance comparison between BAA and OFDMA.}
  \label{Fig:analog_vs_digital}
    \vspace{-4mm}
\end{figure}

\vspace{-4mm}
\subsection{Performance Comparison: Broadband Analog Aggregation v.s. Broadband Digital Aggregation}
The learning accuracy and communication latency of the BAA and the digital OFDMA are compared in Fig. \ref{Fig:analog_vs_digital} under the same transmit SNR per user and a fixed user scheduling scheme with $\frac{R_{\sf in}}{R} = 0.5$. For the digital OFDMA, model-update parameters are quantized into bit sequence with 16-bit per parameter, and the adaptive MQAM modulation is used to maximize the spectrum efficiency under a target BER of $10^{-3}$. 
As shown at Fig. \ref{subfig:accuracy}, although BAA is expected to be more vulnerable to channel noise, it is interesting to see that the two schemes are comparable in learning accuracy (for both the IID and non-IID settings). Such accurate learning of BAA is partly due to the high expressiveness of the deep neural network which makes the learnt model robust against perturbation by channel noise. The result has a profound and refreshing implication that reliable communication may not be the primary concern in edge learning. Essentially, BAA exploits this relaxation on communication reliability to trade for a low communication latency as shown at Fig. \ref{subfig:latency}. The latency gap between the two schemes is remarkable. Without compromising the learning accuracy, BAA can achieve a significant latency reduction ranging from 10x to 1000x. In general, the superiority in latency of BAA over OFDMA is more pronounced in the low SNR regime and dense-network scenarios.

\vspace{-4mm}
\section{Extensions and Discussion}
%In this section, several possible extension directions of the proposed scheme will be briefly discussed.

\vspace{-1mm}
\subsection{Robust Design Against Adversarial Attacks by Spread Spectrum}
One practical issue to be concerned in the federated edge learning is the vulnerability to the adversarial attack by some hostile users who purposely upload inaccurate model-updates or random noise during the model aggregation. The attack may lead to the divergence of the learning algorithm and thus crash down the whole training process. This motivates the design of robust BAA that can suppress the attack from adversarial users as discussed in the sequel. 

\begin{figure}[tt]
\centering
\includegraphics[width=14cm]{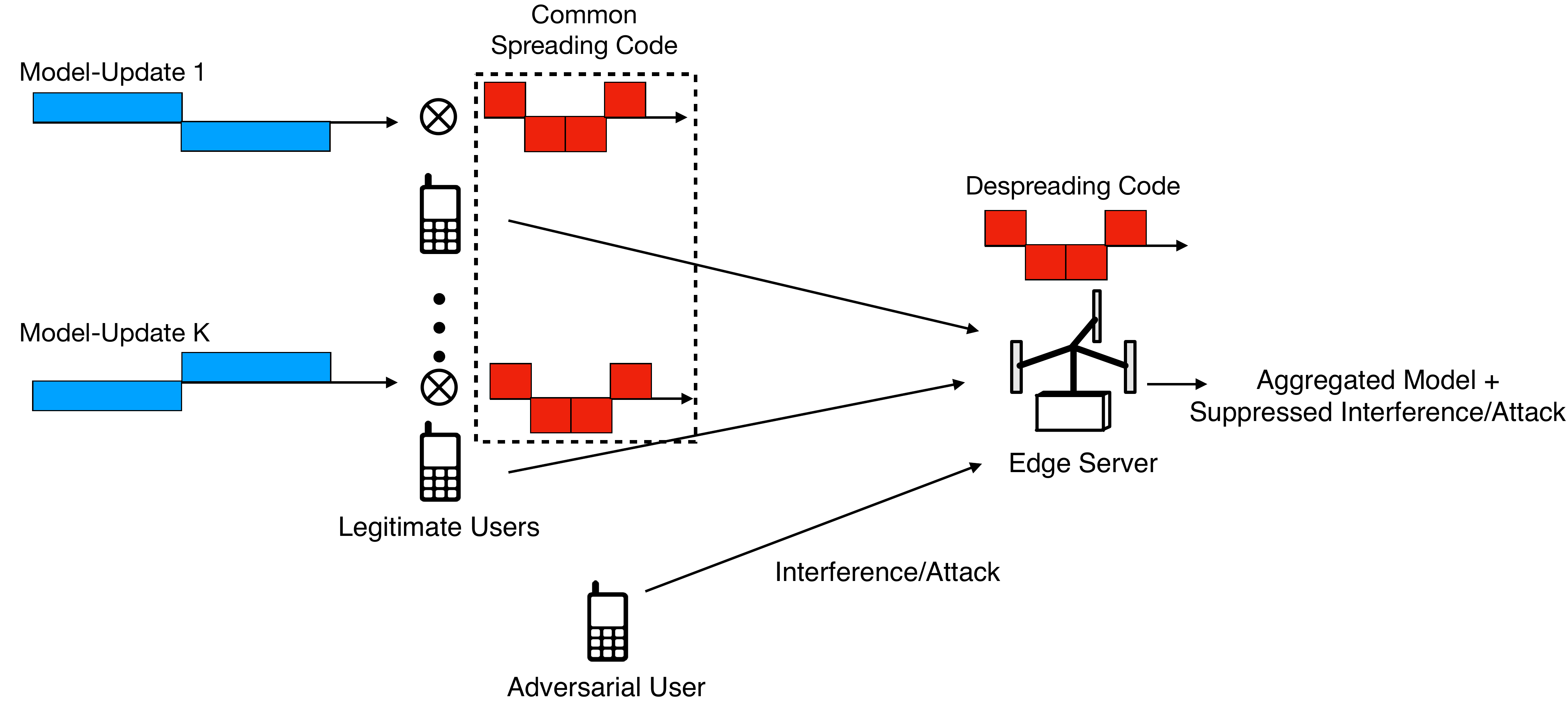}
\caption{Illustration of robust BAA by direct sequence spread spectrum technique.}
\label{Fig:DSSS_AirComp_diagram}
\vspace{-4mm}
\end{figure}

The proposed solution is to deploy the  \emph{direct sequence spread spectrum} (DSSS) technique \cite{madhow1994mmse} to encode the model-updates before transmission. As illustrated in Fig. \ref{Fig:DSSS_AirComp_diagram}, the basic idea of the design is to ensure all legitimate users to use a \emph{common spreading code} assigned by the server to facilitate protected model aggregation. While the adversarial user who is not aware of the spreading code can have its interference suppressed in the despreading/decoding process at the server. The signalling protocol of the design is summarized as follows.

%\begin{framed}
%\vspace{-10pt} 
%\begin{protocol}
\noindent \underline{\bf DSSS-based BAA}:
\begin{enumerate}
\item All legitimate devices in the network are assigned by the server a common spreading code, i.e., a \emph{pseudorandom-noise} code sequence taking value $1$ or $-1$, unknown to the adversarial device. 
\item Before transmission, all legitimate devices will scramble its model-update symbols with the assigned spreading code. 
\item Last, at the edge server side, the received superimposed signal containing all simultaneous transmitting signals will be despread using the known legitimate spreading code. 
\end{enumerate}
%\end{protocol}
%\vspace{-10pt} 
%\end{framed}

\begin{figure}[tt]
\centering
\includegraphics[width=14cm]{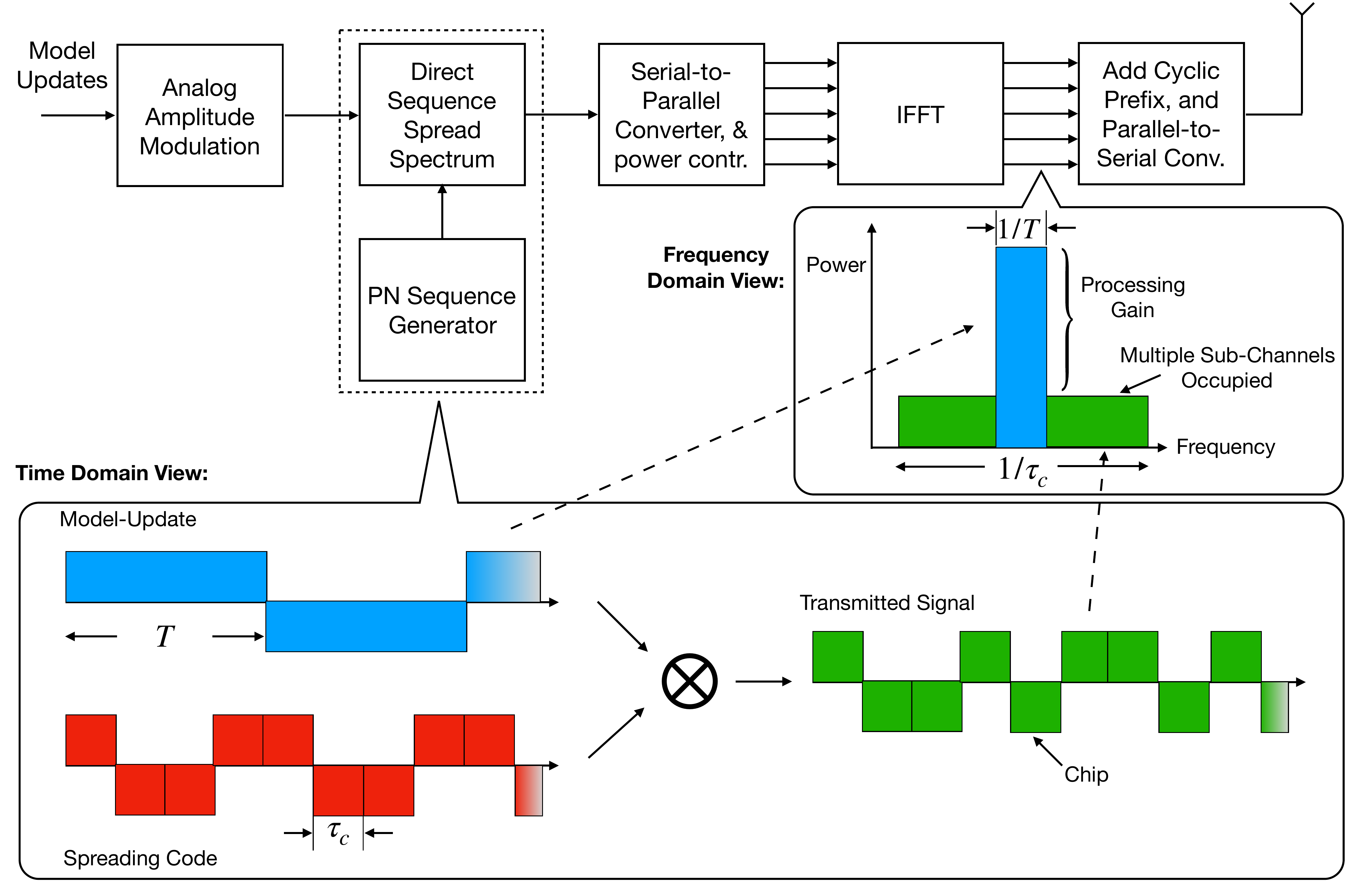}
\caption{System architecture of DSSS-based BAA.}
\label{Fig:DSSS_AirComp_system_architecture}
\vspace{-4mm}
\end{figure}

Since all legitimate devices share the common speading code, the de-speading operation on the received superimposed  signal at the edge server can automatically yield the desired aggregation of the original model-updates from the legitimate devices while suppressing that from the adversarial device by the spectrum-spreading factor defined shortly. 

% In general, the longer spreading code sequence used the better suppression can be made to the adversarial attach, since the correlation between two arbitrary sequence dramatically decreases as sequence dimension grows.

%The system architecture implementing the idea is depicted in Fig. \ref{Fig:DSSS_AirComp_system_architecture}. 
%Compared with the original architecture shown in Fig. \ref{subfig:tx}, 

As shown in Fig. \ref{Fig:DSSS_AirComp_system_architecture}, the implementation of DSSS-based BAA simply involves adding an additional block between the analog amplitude modulation and the IFFT block in the original architecture shown in Fig. \ref{subfig:tx}. 
%It is well-known in the classic DSSS design that, 
It is also illustrated  in Fig. \ref{Fig:DSSS_AirComp_system_architecture} that the multiplication between the model-update with a high-rate spreading code is equal to spreading the bandwidth used for transmitting the update by a factor of 
\begin{align}\label{eq:spreading_factor}  (\text{Spreading factor}) \qquad
\gamma = \frac{T}{\tau_c}, 
\end{align}
where $T$ denotes the symbol duration of the original model-update and $\tau_c$ the duration of a \emph{chip}. 

Therefore, the spreading factor in \eqref{eq:spreading_factor} controls a tradeoff between the latency-reduction ratio of the BAA scheme and the enhancement in receive SNR of model-updates as elaborated below.
\begin{Remark}[Latency-reduction ratio v.s. SNR enhancement]\emph{
As mentioned, by using DSSS technique, the system consumes $\gamma$-times more bandwidth for  model-update transmission. This, to some extent, compromise the latency-reduction ratio achieved by the BAA over the digital counterpart. Nevertheless, the cost is compensated  by safety as well as an improved SNR by the spreading factor. 
}
\end{Remark}

\vspace{-3mm}
\subsection{Coping with Cell-Edge Devices by Beamforming}
\vspace{-1mm}
As mentioned, the update reliability of the proposed BAA scheme is limited by cell-edge devices. The bottleneck can be alleviated by beamforming  if a multi-antenna array is available at the edge server. The key idea is to form sharp beams towards those cell-edge devices to compensate their path-loss so that the receive SNR in \eqref{aligned_power} limited by the furthest devices can be improved.  

The resultant aggregation-beamforming for cell-edge device enhancement differs from the conventional \emph{space division multiple access} (SDMA) beamforming in design principle. Essentially, aggregation-beamforming aims at \emph{amplitude alignment between different received signals}, while SDMA attempts to suppress multi-user interference so as to recover individual data streams from different devices. The difference can be crystalized via their problem formulation as follows. 

 \begin{figure*}[tt]
  \centering
  \subfigure[Aggregation-beamforming]{\label{subfig:AirComp_beam}\includegraphics[width=0.38\textwidth]{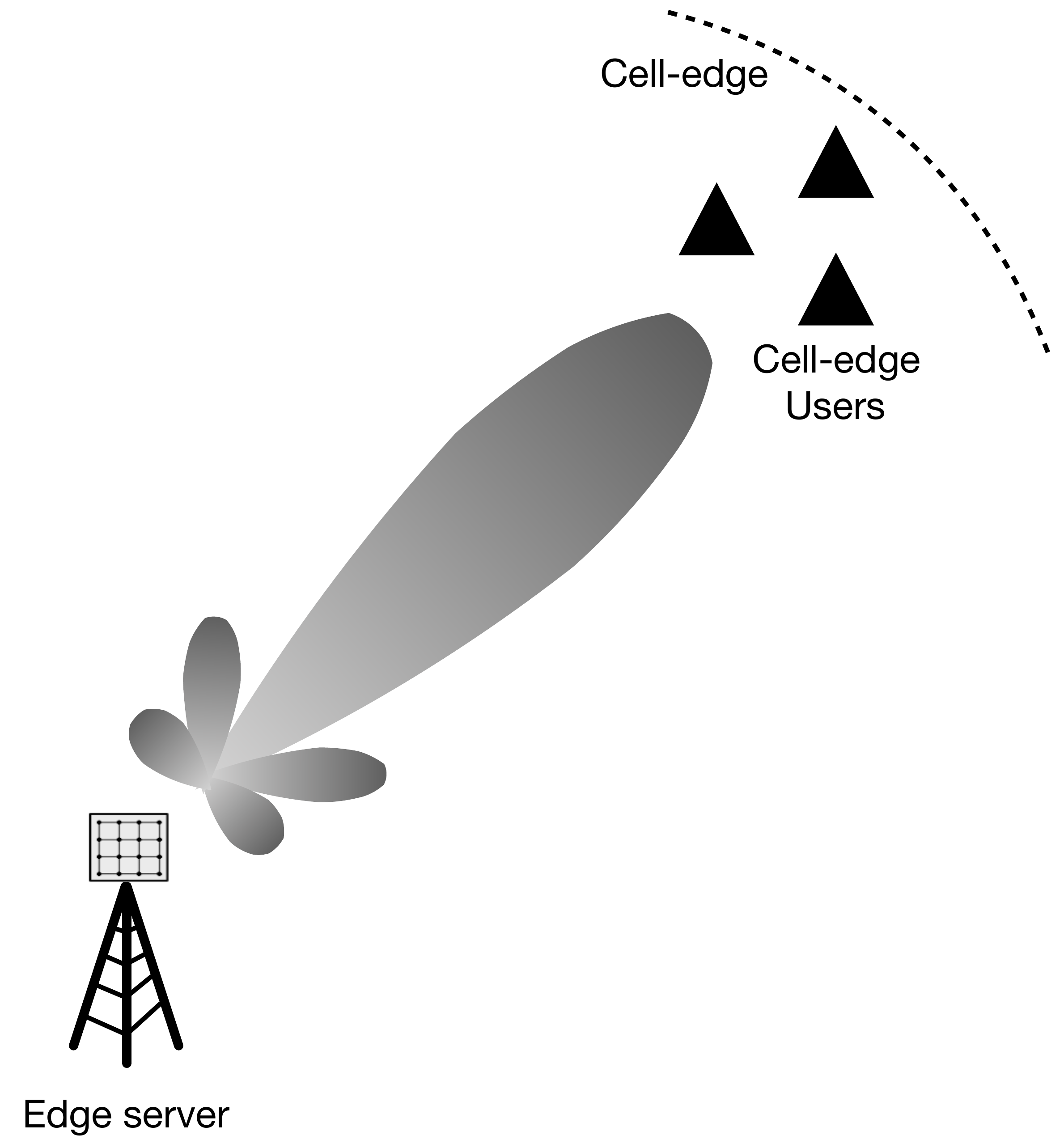}}
  \hspace{0.35in}
  \subfigure[SDMA-beamforming]{\label{subfig:SDMA}\includegraphics[width=0.35\textwidth]{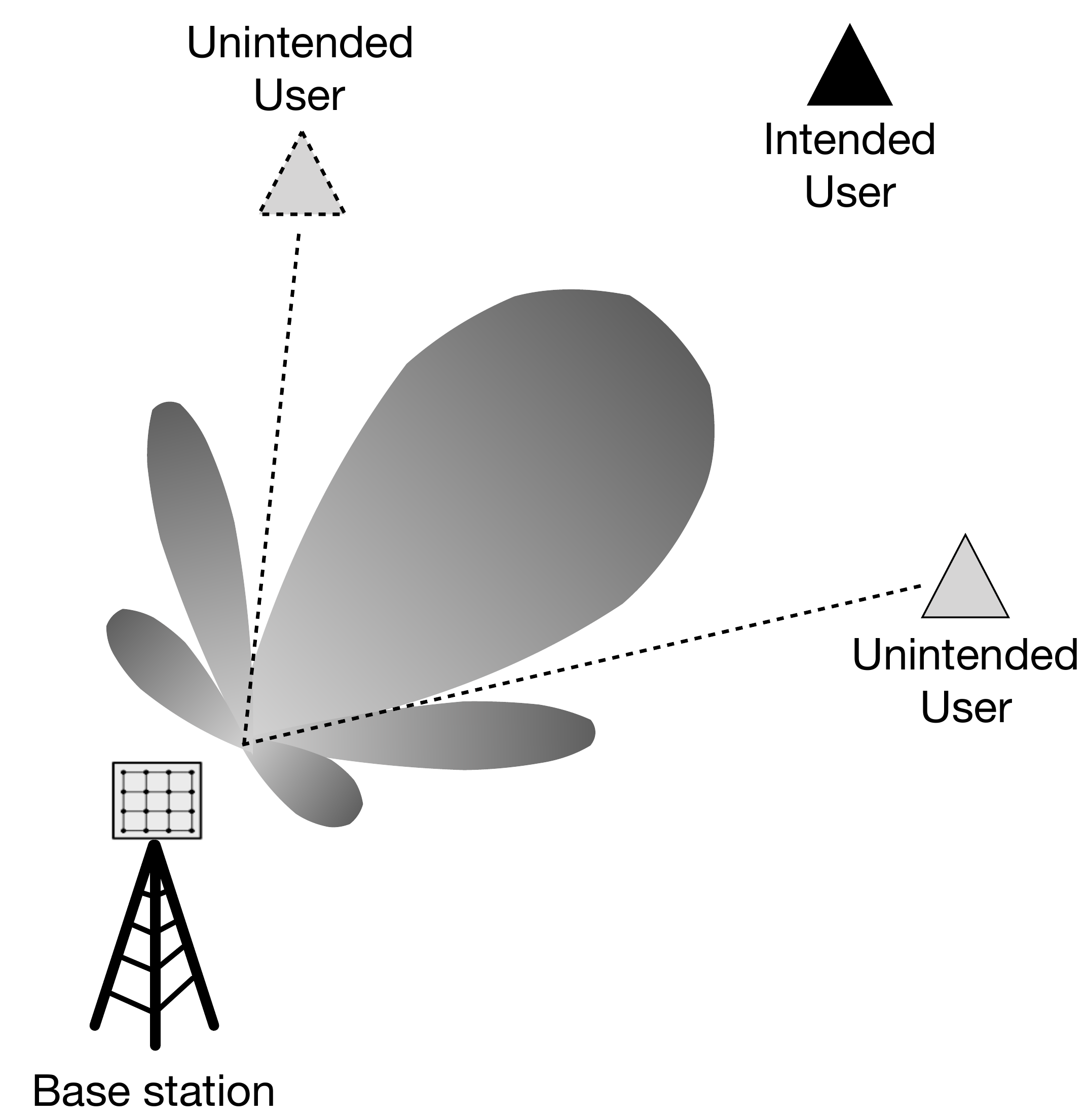}}
  \vspace{3mm}
  \caption{Illustration of aggregation-beamforming and SDMA-beamforming.}
  \label{Fig:AirComp_beam_vs_SDMA}
  \vspace{-3mm}
\end{figure*}

To this end, we consider the following multi-antenna system model:
\begin{align}\label{MIMO_channel}
\by = \bF^H \bH \bx + \bF^H \bn,
\end{align}
where $\bF \in \mathbb{C}^{N \times K}$ is the beamforming matrix to be designed and $N$ denotes the number of antennas equipped at the edge server; $\bH \in \mathbb{C}^{N \times K}$ is the channel matrix whose $k$-th column represents the channel vector for the $k$-th device. $\bx \in \mathbb{C}^K$ is the transmitted symbol vector with the $k$-th element being the symbol from the device $k$. $\bn$ is the AWGN vector with ${\sf E}(\bn \bn^H) = N_0 \bI$.
According to \eqref{MIMO_channel}, the aggregation-beamforming for cell-edge device enhancement can be designed by solving the following \emph{unconstrained SNR maximization} problem: 
\begin{align}\label{AirComp_beam}  (\text{Aggregation-beamforming}) \qquad
\max_{\bF} \frac{{\sf Tr}\l( \bF^H \tilde \bH \tilde \bH^H \bF \r)}{N_0 {\sf Tr}(\bF^H \bF)},
\end{align}
where $\tilde \bH$ contains the channel vectors of the weak users to be enhanced.
On the other hand, in SDMA, each column of $\bF$, denoted by ${\bf f}_k$, represents a dedicated beam targeting an intended user while nulling the interference from others. It thereby yields the following \emph{constrained SNR-maximization} problem. 
\begin{align}\label{SDMA_beam}       (\text{SDMA-beamforming}) \qquad
\begin{aligned}
\max_{{\bf f}_k} \;\; &\frac{{\bf f}_k^H \bh_k}{\sum_{g \neq k} {\bf f}_k^H \bh_g + N_0} \qquad \forall k \\
\text{s.t.} \;\; & {\bf f}_k^H \bh_g = 0, \qquad  \forall g\neq k.
\end{aligned}
\end{align}

\begin{Remark}[Feasibility condition]\label{remark:Feasibility_condition} \emph{
A comparison between the problem in \eqref{AirComp_beam} and that in \eqref{SDMA_beam} reveals that, the implementation of SDMA requires $N \geq K$ to ensure there are sufficient DoFs to enforce the zero-forcing constraints. 
This may not be feasible for the large-scale network with a large $K$, e.g., $100-1000$. In contrast, the aggregation-beamforming is always feasible while more DoFs can lead to a higher SNR enhancement to the weak users. }
\end{Remark}

\begin{Remark}[Beam pattern comparision]\label{remark:Beam_pattern} \emph{
The resultant beamformer patterns from the two formulations are compared in Fig.\ref{Fig:AirComp_beam_vs_SDMA}. In general, aggregation-beamforming can form sharper and stronger beams towards the targeted cell-edge users as the full DoFs are used for SNR enhancement. On the contrary, the beams formed by SDMA towards the intended users tend to be flatter and weaker as the  interference nulling constraints consume part of the DoFs, leaving only a fraction of DoFs for SNR enhancement. Furthermore, users near to each other may cause the differentiability issue in SDMA, due to a finite spatial resolution. This, however, is irrelevant to aggregation-beamforming as no discrimination between users is required. }
\end{Remark}

\section{Concluding Remarks}
In this paper, we have presented the framework of BAA for low-latency FEEL. The design exploits 
the wave-from superposition property of a multi-access channel for communication-efficient update aggregation.
The significance of the work lies in the finding of two communication-and-learning tradeoffs, namely the \emph{SNR-truncation tradeoff} resulting from the amplitude alignment required for aggregation, and the \emph{reliabilty-quantity tradeoff} due to the scheduling of cell-interior devices for constraining path loss. The tradeoffs are fundamental for FEEL network with BAA and can  provide useful guidelines for network planning and optimization.  Besides the findings in tradeoffs, we also prove that the latency-reduction ratio of the proposed BAA w.r.t. the traditional OFDMA scheme scale almost linearly with the device population, justifying the claimed low-latency property.

% To overcome channel fading and achieve the amplitude alignment required for aggregation, the power control polity based on truncated channel inversion is adopted.
% To compensate the undesired channel weighting on the model-updates, truncated channel inversion power adaptation policy is employed. Particularly, to align the receive power for the cell-centric and cell-edge users, a user scheduling problem, with a unique tradeoff between the resultant receive SNR and the data-exploitation ratio during learning, is identified.  An cell-interior scheduling solution giving a flexible control of the above tradeoff is proposed and the performance of which is characterized.  Furthermore, the latency-reduction ratio of the developed BAA solution of the conventional orthogonal multiple access scheme is also derived in closed-form, justifying the claimed low-latency property.  

At a higher lever, the current work represents an initial but important steps towards the fusion of communication and computation/learning. It opens several directions for further investigation. One direction is to further enhance the 
aggregation performance of BAA by exploiting the clustering structure in device distribution for scheduling. Another interesting direction is to integrate the BAA design with the sparsity-induced update-compression techniques for further reducing the communication overhead. Last, robust BAA against the synchronization error is also an important topic to be addressed.

% Another interesting direction is to apply the broadband AirComp design to other IoT-related applications such as high-mobility UAV networks or cloud coordinated vehicular platooning.

%
%XXX: Analog joint source and channel coding for enhancing the communication reliability. 
%
%XXX: use the spreading code as the key and all the participating users share the same key but not the adversarial user. 
%
%XXX: despreading the code can also enhance SNR due to the processing gain attained. 

\appendix
\vspace{-2mm}
\subsection{Proof of Lemma \ref{lemma:1}} \label{app:lemma:1}
\vspace{-2mm}
By definition,
we can establish the following event equivalence. 
\begin{align}\label{eq:equiv_event}
{\sf Pr}(K_{\sf in} = k) = {\sf Pr}\l(k\; \text{devices lie in the range of}\; R_{\sf in}\; \text{while} \; (K-k)\; \text{ones out of the range of} \; R_{\sf in}\r).
\end{align}

Since the device-locations are i.i.d. distributed, the events defined on the right hand side in \eqref{eq:equiv_event} follows a Binomial distribution with the success probability equal to ${\sf Pr}(r_k \leq R_{\sf in})$, i.e., the probability that a device lie in the range of $R_{\sf in}$:
\begin{align}\label{eq:binom}
{\sf Pr}(K_{\sf in} = k) = \binom{K}{k} \l[{\sf Pr}(r_k \leq R_{\sf in})\r]^k \; \l[1 - {\sf Pr} (r_k \leq R_{\sf in})\r]^{K-k}.
\end{align}

Then, according to the uniform distribution presented in \eqref{pdf_r_k}, we have
\begin{align}\label{eq:PDF_r_k}
{\sf Pr}(r_k \leq R_{\sf in}) = \int_0^{R_{\sf in}}  \frac{2r}{R^2} dr = \frac{R_{\sf in}^2}{R^2}.
\end{align}
Thereby, by substituting \eqref{eq:PDF_r_k} into \eqref{eq:PDF_r_k}, the desired result is obtained.

\vspace{-2mm}
\subsection{Proof of Lemma \ref{prop:3}} \label{app:prop:3}
\vspace{-2mm}
By using Lemma \ref{lemma:2} and \eqref{aligned_power}, the expected receive SNR of all-inclusive scheme can be computed by
\begin{align}\label{app:prop:3:eq:1}
{\sf E}(\rho_0) &= \int_0^R   \frac{P_0}{M x^\alpha {\sf Ei}(g_{\sf th})} f_{r_{\max}}(x)  dx \notag\\
& =  \frac{P_0}{M {\sf Ei}(g_{\sf th})} \frac{2K}{R^{2K}} \int_0^R x^{2K - \alpha -1} dx
\end{align}

To ensure that the integral in \eqref{app:prop:3:eq:1} converges, it requires that $2K - \alpha -1 \geq 0$. The assumption always holds in practice as mentioned earlier. Under the assumption for convergence, by completing the integral, we can have the desired result:
\begin{align}
{\sf E}(\rho_0) = \frac{2K}{2K-\alpha} \frac{P_0}{M R^\alpha {\sf Ei}(g_{\sf th})}.
\end{align}

\vspace{-2mm}
\subsection{Proof of Lemma \ref{prop:4}} \label{app:prop:4}
\vspace{-2mm}
For the cell-interior scheduling, the expectation on the  receive SNR is more challenging to derive, as the number of scheduled devices is now a random variable, adding an additional layer of randomness to the receive SNR besides the randomly  distributed device-distance. 

To overcome the challenge, we find it convenient to tackle the two-layer randomness sequentially using the trick of conditional expectation. Particularly, the expected aligned received power can be computed using the following formula. 
\begin{align}\label{eq:two_layered_expectation}
{\sf E}(\rho_0) = {\sf E}[{\sf E}(\rho_0 \mid K_{\sf in} = k)],
\end{align}
\vspace{-2mm}
where the first expectation is taken over the $k$-th furthest distance to the edge server conditioned on $K_{\sf in} = k$, while the second expectation is over the variable $K_{\sf in}$ whose PMF is given in Lemma \ref{lemma:1}.
Then \eqref{eq:two_layered_expectation} can be explicitly written as 
\begin{align}\label{eq:two_layered_expectation2}
{\sf E}(\rho_0) = \sum_{k=0}^K  {\sf E}(\rho_0 \mid K_{\sf in} = k) {\sf Pr}(K_{\sf in} = k).
\end{align}

For simplicity, we consider the typical case that $\alpha = 3$. Note that the first term in \eqref{eq:two_layered_expectation} is equal to zero, i.e., ${\sf E}(\rho_0 \mid K_{\sf in} = 0) = 0$, and the second term is negligible when $K$ is sufficiently large since ${\sf Pr}(K_{\sf in} = 1) \to 0$ and ${\sf E}(\rho_0 \mid K_{\sf in} = k)$ should be bounded.

Then remaining task is to compute ${\sf E}(\rho_0 \mid K_{\sf in} = k)$ for $k \geq 2$. Note that given $K_{\sf in} = k$, the $k$ scheduled devices also follow i.i.d. uniform distribution over the cell-interior within the distance of $R_{\sf in}$. 
As a result, by following simular steps in the proof of Lemma \ref{prop:3}, for $k \geq 2$,
one can easily derive that, 
\begin{align}\label{eq:conditioned_expectation}
{\sf E}(\rho_0 \mid K_{\sf in} = k) = \frac{2k}{2k-\alpha} \frac{P_0}{M R_{\sf in}^\alpha {\sf Ei}(g_{\sf th})},
\end{align}

Substituting \eqref{eq:conditioned_expectation} into \eqref{eq:two_layered_expectation2}, it follows that
\begin{align}
{\sf E}(\rho_0) = \frac{P_0}{M R_{\sf in}^\alpha {\sf Ei}(g_{\sf th})} \underbrace{\sum_{k = 2}^K \frac{2k}{2k - \alpha} \binom{K}{k} \l(\frac{R_{\sf in}^2}{R^2} \r)^k \l( 1-\frac{R_{\sf in}^2}{R^2} \r)^{K-k}}_{c(R_{\sf in})},
\end{align}
which gives the derived result in \eqref{eq:prop:4}.

Also note that the scaling factor $c(R_{\sf in})$ is essentially a weighted average for the term $\frac{2k}{2k - \alpha}$ from $k=2$ to $K$. Given that $\alpha =3$, and $K$ is sufficiently large, we note that $\frac{2k}{2k - \alpha}$ monotonically ranges from $1$ to $4$. Since a weighted average for the values from a range will not exceed the range, it gives the conclusion that   $1 \leq c(R_{\sf in}) \leq 4$, which completes the proof.

\bibliographystyle{ieeetr}
\bibliography{BibDesk_File}

\end{document}